\documentclass[a4paper,11pt]{article}
\pdfoutput=1 

\usepackage{jcappub} 

\def\doi{http://doi.org}

\newtheorem{remark}{Remark}[section]

\newcommand{\HCd}{\mathcal{H}}

\def\HCdt0{\tilde{\HCd}_{0}}

\usepackage[T1]{fontenc} 

\title{ Quintessential Inflation and Cosmological Seesaw Mechanism:\\ Reheating and Observational Constraints}

\author[a]{L. Arest\'e Sal\'o,}
\author[b, c, d]{D. Benisty,}
\author[b, d, e]{E. I. Guendelman,}
\author[f,2]{J. d. Haro}

\affiliation[a]{School of Mathematical Sciences, Queen Mary University of London, Mile End Road, London, E1 4NS, United Kingdom}
\affiliation[b]{Physics Department, Ben-Gurion University of the Negev, Beer-Sheva 84105, Israel}
\affiliation[c]{DAMTP, Centre for Mathematical Sciences, University of Cambridge, Wilberforce Road, Cambridge CB3 0WA, United Kingdom}

\affiliation[d]{Frankfurt Institute for Advanced Studies (FIAS), Ruth-Moufang-Strasse~1, 60438 Frankfurt am Main, Germany}

\affiliation[e]{Bahamas Advanced Study Institute and Conferences, 4A Ocean Heights, Hill View Circle, Stella Maris, Long Island, The Bahamas}

\affiliation[f]{Departament de Matem\`atiques, Universitat Polit\`ecnica de Catalunya, Diagonal 647, 08028 Barcelona, Spain}

\emailAdd{l.arestesalo@qmul.ac.uk}
\emailAdd{benidav@post.bgu.ac.il}
\emailAdd{guendel@bgu.ac.il}
\emailAdd{jaime.haro@upc.edu}

\abstract{Recently a new kind of quintessential inflation coming from the Lorentzian distribution has been introduced in \cite{Benisty:2020xqm,Benisty:2020qta}. The model leads to a very simple potential, which basically depends on two parameters, belonging to the class of $\alpha$-attractors  and depicting correctly the early and late time accelerations of our universe. The potential emphasizes a {\it cosmological seesaw mechanism} (CSSM) that produces a large inflationary vacuum energy in one side of the potential and a very small value of dark energy on the right hand side of the potential. {Here we show that the model agrees with the recent observations and with the reheating constraints. Therefore the model gives a reasonable scenario beyond the standard $\Lambda$CDM that includes the inflationary epoch.}}

\begin{document}
\maketitle
\flushbottom

\section{Introduction}
After the discovery of the current cosmic acceleration \cite{Weinberg:1988cp,Lombriser:2019jia,Frieman:2008sn,Riess:2019cxk}, several theoretical mechanisms were developed in order to explain it. One of them is quintessence (see for instance \cite{Caldwell:1997ii,Ratra:1987rm,Peebles:1987ek,Barreiro:1999zs,Carroll:1998zi,Chiba:1999wt,Sahni:1999qe,Krishnan:2020vaf,Krishnan:2021dyb}),
where a scalar field is the responsible for the late time acceleration of our universe. The next step was to unify both acceleration phases: the early acceleration of the universe, named {\it inflation} \cite{Guth:1980zm,Linde:1981mu,Starobinsky:1980te}, with the current acceleration. One of the simplest ways to do it is the so-called {\it quintessential inflation}, introduced for the first time by Peebles and Vilenkin in \cite{Peebles:1998qn}, where the inflaton field is the only responsible for both inflationary phases.

Several authors developed and improved the original Peebles-Vilenkin model \cite{Dimopoulos:2001ix, Dimopoulos:2017zvq,Hossain:2014coa,Hossain:2014xha,Hossain:2014zma,Haro:2019gsv,deHaro:2016hpl,Geng:2015fla,Geng:2017mic,deHaro:2016hsh,deHaro:2016ftq,deHaro:2016cdm,deHaro:2017nui, AresteSalo:2017lkv,Haro:2015ljc,Rubio:2017gty,Dimopoulos:2018wfg,Staicova:2018bdy,Haro:2019peq,Guendelman:2014bva,Guendelman:2015liz,vandeBruck:2017voa,Dimopoulos:2019ogl,Kleidis:2019ywv,Lima:2019yyv,Staicova:2020wph,Benisty:2019vej,Rosa:2019jci,Staicova:2019ksr,Dimopoulos:2020pas,Es-haghi:2020oab,Banerjee:2020xcn,Rodrigues:2020dod} obtaining models whose theoretical results match very well with the observational data provided by the Planck's team \cite{Ade:2015xua,Akrami:2018odb,Aghanim:2018eyx}. A simple model, constructed from the well-known Lorentzian distribution, was recently presented in \cite{Benisty:2020xqm,Benisty:2020qta}. 
{  We show that  the model 
provides the same spectral index and ratio of tensor to scalar perturbation as the  $\alpha$-attractors models
\cite{Linde:1981mu,Akrami:2017cir,Akrami:2020zxw,Rodrigues:2020jsv,Elizalde:2015nya,Dubinin:2017irq,Pozdeeva:2020shl},}
meaning that it yields a power spectrum of perturbations agreeing with the observation data and is able to depict correctly the current cosmic acceleration. Another property of the model is that it provides a {\it seesaw mechanism}. In the theory of grand unification of particle physics and, in particular, in theories of neutrino masses and neutrino oscillation, the seesaw mechanism is a generic model used to understand the relative sizes of observed neutrino masses of the order of $eV$, compared to those of quarks and charged leptons, which are millions of times heavier \cite{Minkowski:1977sc,Yanagida:1980xy,Schechter:1980gr,Davidson:1987tr,Davidson:1987mh,Rajpoot:1987ji}. The approach adopted in \cite{Guendelman:1999qt,Guendelman:2001bu,Guendelman:2002js,Guendelman:2014bva} explains the difference between the inflationary vacuum energy density value and the late dark energy density value. As in the case of masses difference in particle physics, here the model predicts that as long as the one epoch has a very low energy density the other one has to have very large energy density. In \cite{Benisty:2020xqm,Benisty:2020qta} a potential is constructed from the Lorentzian form of the $\epsilon$ parameter. Here we consider some scalar potential from the beginning which implements the same features of the original models, with much less parameter numbers. We study in detail this simple model, which only depends on two parameters, and we show its viability.

The paper is organized as follows: In Section \ref{sec:LQIm} we introduce the Lorentzian quintessential model, studying its power spectrum during inflation and providing the theoretical value of the parameters involved in the model. Sections \ref{sec:dynamics} and \ref{sec:numerics} are devoted to the study of all the evolution of the inflaton field, showing that the theoretical results provided by the model agree with the current observational data. Then, in Section \ref{sec:othermodels} we discuss the viability of other similar models and in Section \ref{sec:Data} we use a combination of cosmological probes from different data sets to constrain further our model and verify its viability. Finally, Section \ref{sec:summary} summarises the results. The units used throughout the paper are $\hbar=c=1$ and we denote  the reduced Planck's mass by $M_{pl}\equiv \frac{1}{\sqrt{8\pi G}}\cong 2.44\times 10^{18}$ GeV.

\section{The Lorentzian Quintessential Inflation model}
\label{sec:LQIm}
Based on the Cauchy distribution (Lorentzian in the physics language) in \cite{Benisty:2020xqm,Benisty:2020qta} the ansatz to be considered is the following one, 
\begin{eqnarray}\label{ansatz}
\epsilon(N)=\frac{\xi}{\pi}\frac{\Gamma/2}{N^2+\Gamma^2/4},
\end{eqnarray}
where $\epsilon$ is the main slow-roll parameter and $N$ denotes the number of e-folds. From this ansatz, we can find the {  exact} corresponding potential of the scalar field, {   namely
\begin{eqnarray}\label{original}
\hspace{-0.5cm}V(\varphi)=\lambda M_{pl}^4\exp\left[-\frac{2\xi}{\pi}\arctan\left(\sinh
\left(\gamma\varphi/M_{pl} \right)  \right)\right]\boldsymbol{\cdot} 
\left(1-\frac{2\gamma^2\xi^2}{3\pi^3}\frac{1}{\cosh
\left(\gamma\varphi/M_{pl} \right) } \right),
\end{eqnarray}
where $\lambda$ is a dimensionless parameter and the parameter $\gamma$ is defined by
$$\gamma\equiv \sqrt{\frac{\pi}{\Gamma \xi}}.$$ This potential can be derived by using equations (37) in \cite{Martin:2016iqo}. However, in this work we are going to use a more simplified potential,  keeping the same properties as the original potential but not coming from the ansatz (\ref{ansatz}),
}
\begin{eqnarray}\label{exp}
V(\varphi)=\lambda M_{pl}^4\exp\left[-\frac{2\xi}{\pi}\arctan\left(\sinh\left(\gamma\varphi/M_{pl} \right)  \right)\right].
\end{eqnarray}
 
\begin{figure}[t!]
\begin{center}
\includegraphics[width=15cm]{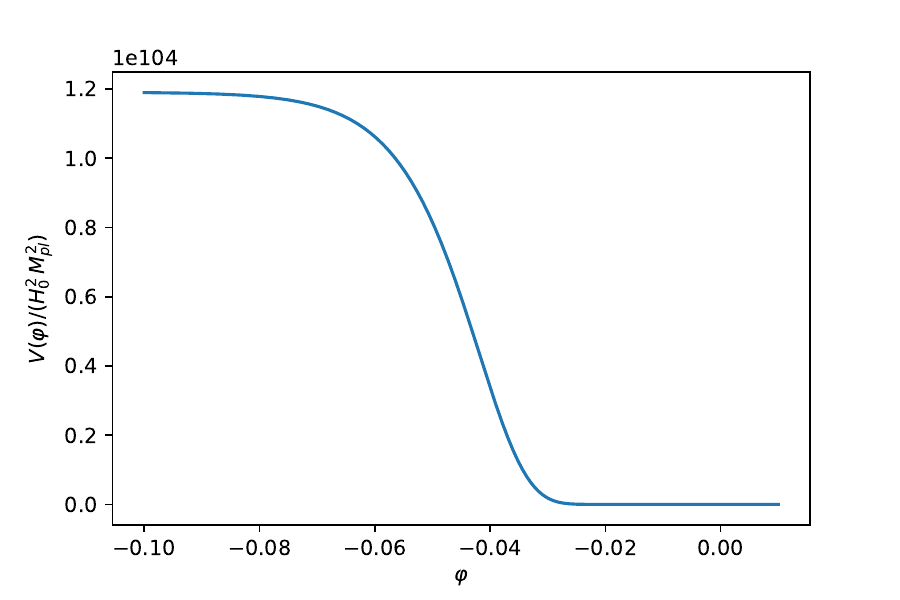}
\caption{\it{The shape of the scalar potential \eqref{exp} with  $\xi\sim 122$ and $\lambda\sim 10^{-69}$. The left side shows the inflationary energy density and the right side shows the late dark energy density. }}
\label{fig:potential}
\end{center}
\end{figure}

We can see the shape of the potential on Fig. \ref{fig:potential}, where the inflationary epoch takes place on the left hand side of the graph, while the dark energy epoch occurs on the right hand side.

The main slow roll parameter is given by 
\begin{eqnarray}
\epsilon\equiv \frac{M_{pl}^2}{2}\left( \frac{V_{\varphi}}{V} \right)^2= \frac{2\xi/(\Gamma\pi)}{\cosh^2\left(\gamma\frac{\varphi}{M_{pl}} \right)}
=\frac{2\gamma^2\xi^2/\pi^2}{\cosh^2\left(\gamma\frac{\varphi}{M_{pl}} \right)}
\end{eqnarray}
and, since inflation ends when $\epsilon_{END}=1$, one has to assume that $\frac{2\xi}{\Gamma\pi}>1$ to guarantee the end of this period.

In fact, we have
\begin{eqnarray}
\varphi_{END}=\frac{M_{pl}}{\gamma}\ln\left( \sqrt{\frac{2\xi}{\Gamma\pi}} -\sqrt{\frac{2\xi}{\Gamma\pi}-1} \right)=\\
\frac{M_{pl}}{\gamma}\ln\left[\frac{\sqrt{2}\xi}{\pi}\left( \gamma-\sqrt{\gamma^2-\frac{\pi^2}{2\xi^2}}  \right)\right]<0\nonumber
\end{eqnarray}
and we can see that, for large values of $\gamma$, one has that $\varphi_{END}$ is close to zero. Thus, we will choose $\gamma\gg 1\Longrightarrow \Gamma\xi\ll 1$, which is completely compatible with the condition $\frac{2\xi}{\Gamma\pi}>1$.

On the other hand, the other important slow roll parameter is given by
\begin{eqnarray}
\eta\equiv M_{pl}^2\frac{V_{\varphi\varphi}}{V}=\frac{2\xi\gamma^2}{\pi}\frac{\tanh\left(\gamma\frac{\varphi}{M_{pl}}  \right)}{\cosh\left(\gamma\frac{\varphi}{M_{pl}} \right)}
+\frac{4\gamma^2\xi^2/\pi^2}{\cosh^2\left(\gamma\frac{\varphi}{M_{pl}} \right)}.\end{eqnarray}
Both slow roll parameters have to be evaluated when the pivot scale leaves the Hubble radius, which will happen for large values of 
$\cosh\left(\gamma\varphi/M_{pl} \right)$, obtaining 
\begin{equation}
\epsilon_*=\frac{2\gamma^2\xi^2/\pi^2}{\cosh^2\left(\gamma\frac{\varphi_*}{M_{pl}}  \right)}, \quad \eta_*\cong \frac{2\xi\gamma^2}{\pi}\frac{\tanh\left(\gamma\frac{\varphi_*}{M_{pl}}  \right)}{\cosh\left(\gamma\frac{\varphi_*}{M_{pl}}  \right)}
\end{equation}
with $\varphi_{*}<0$. Then, since the spectral index is given in the first approximation by $n_s\cong 1-6\epsilon_*+2\eta_*$,  one gets after some algebra
\begin{eqnarray}
n_s\cong 1+ 2\eta_*\cong 1-\gamma\sqrt{r/2},
\end{eqnarray}
where $r=16\epsilon_*$ is the ratio of tensor to scalar perturbations.

Now, we calculate the number of e-folds from the leaving of the pivot scale  to the end of inflation, which is given by
\begin{eqnarray}
N=\frac{1}{M_{pl}}\int_{\varphi_*}^{\varphi_{END}}\frac{1}{\sqrt{2\epsilon}}d\varphi= \frac{\pi}{2\gamma^2\xi}
\left[\sinh\left(\gamma\varphi_{END}/M_{pl} \right)-\sinh\left(\gamma\varphi_*/M_{pl} \right) \right]\cong \frac{\xi}{\sqrt{2\epsilon_*}}\nonumber,
\end{eqnarray}
so we have that
\begin{eqnarray}\label{n_s-r}
n_s\cong 1-\frac{2}{N},\qquad r\cong\frac{8}{N^2\gamma^2},
\end{eqnarray}
meaning that our model {  leads to the same spectral index and tensor/scalar ratio as the  $\alpha$-attractors models} with $\alpha=\frac{2}{3\gamma^2}$  (see for instance \cite{Linde:1981mu}).

\begin{remark}
For the original potential coming from the ansatz (\ref{ansatz}), i.e., for the potential given in the formula (\ref{original}),
one can use the slow roll parameters $\epsilon_1=\epsilon$
and $\epsilon_2=\frac{d\ln \epsilon_1}{dN}=2(2\epsilon-\eta)$ to  obtain, for large values of the number of e-folds,
\begin{eqnarray}
\epsilon_1=\frac{1}{2\gamma^2N^2} \quad \mbox{and} \quad \epsilon_2=-\frac{2}{N},
\end{eqnarray}
and thus, taking into account that $n_s=1-2\epsilon_1-\epsilon_2$, one easily gets the result given in the formula (\ref{n_s-r}).
Unfortunately, since our potential (\ref{exp}) is a simplified version of the original potential coming from the ansatz (\ref{ansatz}), in order to justify the expression of the spectral index and the ratio of tensor to scalar perturbations, we cannot do this simple calculation, which only holds for the original potential, and we must perform all the calculation presented in this section.
\end{remark}

\begin{figure}[t!]
\begin{center}
\includegraphics[width=15cm]{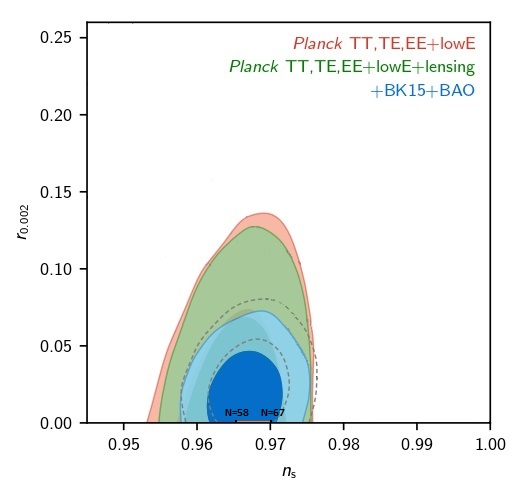}
\caption{\it{The Marginalized joint confidence contours for $(n_s,r)$ at $1\sigma$ and $2\sigma$ CL, without the presence of running of the spectral indices. We have drawn the curve for the present model for $\gamma\gg 1$ from $N=58$ to $N=67$} e-folds. (Figure courtesy of the Planck2018 Collaboration).}
\end{center}
\end{figure}

Finally, it is well-known that the power spectrum of scalar perturbations is given by
\begin{eqnarray}\label{power}
{\mathcal P}_{\zeta}=\frac{H_*^2}{8\pi^2\epsilon_*M_{pl}^2}\sim 2\times 10^{-9}.
\end{eqnarray}

Now, since in our case $V(\varphi_*)\cong \lambda M^4_{pl}e^{\xi}$, meaning that $H_*^2\cong \frac{\lambda M^2_{pl}}{3}e^{\xi}$, and taking into account that
$\epsilon_*\cong \frac{(1-n_s)^2}{8\gamma^2}$, one gets the constraint
\begin{eqnarray}\label{constraint}
\lambda \gamma^2 e^{\xi}\sim 7\times 10^{-11},
\end{eqnarray}
where we have chosen as a value of $n_s$ its central value $0.9649$.

Summing up, we will choose our parameters satisfying the condition (\ref{constraint}), with $\gamma\gg 1$ and $\xi\gg 1$, which will always fulfill the constraints $\Gamma\xi\ll 1$ and $\frac{2\xi}{\Gamma\pi}$ that we have imposed. Then, to find the values of the parameters one can perform the following heuristic argument:

Taking for example $\gamma=10^2$, the constraint  (\ref{constraint}) becomes $\lambda e^{\xi}\sim 7\times 10^{-15}$. On the other hand, at the present time we will have
$\gamma\varphi_0/M_{pl}\gg 1$ where $\varphi_0$ denotes the current value of the field. Thus, we will have $V(\varphi_0)\sim \lambda M_{pl}^4e^{-\xi}$, which is the dark energy at the present time, meaning that 
\begin{eqnarray}
0.7\cong \Omega_{\varphi, 0}\cong \frac{V(\varphi_0)}{3H^2_0 M_{pl}^2}
\sim \frac{\lambda e^{-\xi}}{3}\left(\frac{M_{pl}}{H_0}\right)^2.
\end{eqnarray}
Taking the value $H_0=67.81\; \mbox{Km/sec/Mpc}=5.94\times 10^{-61} M_{pl}$, we get the equations
\begin{eqnarray}
\lambda e^{\xi}\sim 7\times 10^{-15}
\qquad \mbox{and} \qquad \lambda e^{-\xi}\sim 10^{-120},\end{eqnarray}
whose solution is given by  $\xi\sim 122$ and $\lambda\sim 10^{-69}$.

If we choose $\gamma\sim 10^2$,  we see that the values of $\xi$ and $\gamma$ could be set equal in order to obtain the desired results from both the early and late inflation. From now on we will set $\xi=\gamma$, since it may help to find a successful combination of parameters because it reduces the number of effective parameters. As we will see later, numerical calculations show that, in order to have $\Omega_{\varphi,0}\cong 0.7$ (observational data show that, at the present time, the ratio of the energy density of the scalar field to the critical one is approximately $0.7$), one has to choose $\xi=\gamma\cong 121.8$. 

\

To end this section we aim to find the relation between the number of e-folds and the reheating temperature, namely $T_{rh}$. {  For this derivation, we are going to use the same procedure as in \cite{AresteSalo:2017lkv}, which resembles the ones in \cite{Martin:2010kz} and \cite{Martin:2014nya}.}
We start with the formula 
\begin{eqnarray}\frac{k_*}{a_0H_0}=e^{-N}\frac{H_*}{H_0}\frac{a_{END}}{a_{kin}}\frac{a_{kin}}{a_{rh}}
\frac{a_{rh}}{a_{matt}}\frac{a_{matt}}{a_{0}},
\end{eqnarray}
where $a$ is the scale factor and $a_{END}$, $a_{kin}$, $a_{rh}$, $a_{matt}$ and $a_0$ denote respectively its value at the end of inflation, at the beginning of kination, radiation and  matter domination,  and finally at the present time.

Taking into account that 
\begin{eqnarray}
\left(\frac{a_{kin}}{a_{rh}} \right)^6=\frac{\rho_{rh}}{\rho_{kin}} \quad \mbox{and}\quad 
\left(\frac{a_{rh}}{a_{matt}} \right)^4=\frac{\rho_{matt}}{\rho_{rh}},
\end{eqnarray}
and noting that $H_0\sim 2\times 10^{-4} \mbox{Mpc}^{-1}\sim 6\times 10^{-61} M_{pl}$ and $k_*=a_0k_{phys}$, where we have chosen $k_{phys}=0.02\mbox{Mpc}^{-1}$,  we have obtained
\begin{eqnarray}
N=-4.61+\ln\left(\frac{H_*}{H_0}  \right)+ \ln\left(\frac{a_{END}}{a_{kin}}  \right)+\frac{1}{4}\ln\left(\frac{g_{matt}}{g_{rh}}  \right)
+\frac{1}{6}\ln\left(\frac{\rho_{rh}}{\rho_{kin}}  \right)+ \ln\left(\frac{T_0}{T_{rh}}  \right),
\end{eqnarray}
where we have used that after the matter-radiation equality the evolution is adiabatic, that is, $a_0T_0=a_{matt} T_{matt}$ as well as the relations $\rho_{matt}=\frac{\pi^2}{30}g_{matt}T_{matt}^4$ and $\rho_{rh}=\frac{\pi^2}{30}g_{rh}T_{rh}^4$ being $g_{matt}=3.36$ the degrees of freedom at the matter-radiation equality and we have chosen as degrees of freedom at the reheating time the ones of the Standard Model, i.e., 
$g_{rh}=106.75$.

Now, from the formula of the power spectrum (\ref{power}) we infer that $H_*\sim 4\times 10^{-4}\sqrt{\epsilon_*} M_{pl}$, obtaining
\begin{eqnarray}
N=125.37+\frac{1}{2}\ln\epsilon_*+ \ln\left(\frac{a_{END}}{a_{kin}}  \right)
+\frac{1}{6}\ln\left(\frac{\rho_{rh}}{\rho_{kin}}  \right)+ \ln\left(\frac{T_0}{T_{rh}}  \right),
\end{eqnarray}
and introducing the current value of the temperature of the universe $T_0\sim 9.6\times 10^{-32}M_{pl}$ we get 
\begin{eqnarray}
N=54.36+\frac{1}{2}\ln\epsilon_*+ \ln\left(\frac{a_{END}}{a_{kin}}  \right)
-\frac{1}{3}\ln\left(\frac{H_{kin}T_{rh}}{M_{pl}^2 } \right).
\end{eqnarray}

As we will see in next section we have numerically checked that $H_{kin}\sim 4\times 10^{-8} M_{pl}$, which leads to 
\begin{eqnarray}
N+\ln N=54.82
-\frac{1}{3}\ln\left(\frac{T_{rh}}{M_{pl} } \right),
\end{eqnarray}
where we have used that $\epsilon_*=\frac{1}{2\gamma^2N^2}$ and we have also numerically computed that $\ln\left(\frac{a_{END}}{a_{kin}} \right)\cong -0.068$.

Since the scale of  nucleosynthesis is $1$ MeV and in order to avoid the late time decay of gravitational relic products such as moduli fields or gravitinos  which could jeopardise  the nucleosynthesis success, one needs temperatures lower than $10^9$ GeV. So, we will assume that $1 \mbox{MeV}\leq T_{rh}\leq 10^9 \mbox{GeV}$, which leads to constrain the number of e-folds to $58\lesssim N\lesssim 67$. And for this number of e-folds, $0.966\lesssim n_s\lesssim 0.970$, which enters within its $2\sigma$ CL range.

\section{Dynamical evolution of the scalar field}
\label{sec:dynamics}

In this section,  we want to calculate the value of the scalar field and its derivative. In this model, as always happens in quintessential inflation, the early inflation is followed up by a kination phase, which is essential to
match the model with the Hot Big Bang. Effectively, immediately after the end of inflation the potential is so low that the kinetic energy density of the inflaton field dominates, that is, the universe enters in a kination phase, which is characterised  by an effective Equation of State (EoS) parameter $w_{eff}$ equal to $1$ because the potential is negligible. Thus, the energy density of the scalar field decreases as $a^{-6}$, being $a$ the scale factor. On the contrary, the particles produced during the phase transition  between inflation and kination, whose energy density decreases as $a^{-4}$, will eventually dominate and the universe  will become reheated. 

\begin{remark}
Note that this kination phase is not needed in standard inflation where the inflaton field loses all its energy oscillating in the deep well of the potential and producing the particles that will reheat the universe. 
\end{remark}



Then, taking into account the importance of the kination phase in quintessential inflation,
analytical calculations can be done disregarding the potential during kination because during this epoch the potential energy of the field is negligible. Then, since during kination one has $a\propto t^{1/3}\Longrightarrow H=\frac{1}{3t}$, using the Friedmann equation the dynamics in this regime will be
\begin{eqnarray}
\frac{\dot{\varphi}^2}{2}=\frac{M_{pl}^2}{3t^2}\Longrightarrow \dot{\varphi}=\sqrt{\frac{2}{3}}\frac{M_{pl}}{t}\Longrightarrow \\
\varphi(t)=\varphi_{kin}+\sqrt{\frac{2}{3}}M_{pl}\ln \left( \frac{t}{t_{kin}} \right)\nonumber,\end{eqnarray}
where we use by definition \cite{Joyce:1996cp,Spokoiny:1993kt}  as the beginning of the kination the moment
when the Equation of State parameter is close to $1$, which coincides when 
the derivative of the field is maximum, corresponding to $\varphi_{kin}\approx -0.03M_{pl}$ and $w_{\varphi}\approx 0.99$ for $\gamma=\xi=122$.

Recall that for our choice of the  parameters $\varphi_{END}$ is very close to zero and, looking at the shape of the potential, this regime has to start very near from
$\varphi=0$. In order to check it numerically, we have integrated the dynamical system
$$\ddot{\varphi}+3H\dot{\varphi}+V_{\varphi}=0,
$$
with initial conditions when the pivot scale leaves the Hubble radius, that is,
with $\varphi_i=\varphi_*$ and $\dot{\varphi}_i=0$, where
$$
\frac{(1-n_s)^2}{8\gamma^2}=\frac{2\gamma^2\xi^2}{\pi^2}\frac{1}{\cosh^2\left(\gamma\varphi_*/M_{pl} \right)},\qquad n_s=0.9649.$$
Thus, at the reheating time, i.e., at the beginning of the radiation phase, one has 
\begin{eqnarray}
\varphi_{rh}=\varphi_{kin}+\sqrt{\frac{2}{3}}M_{pl}\ln\left( \frac{H_{kin}}{H_{rh}} \right),
\end{eqnarray}
where we assume, as usual, that there is not drop of energy from the end of inflation to the beginning of kination, i.e.,
$H_{kin}= H_{END}=\frac{\sqrt{V(\varphi_{END})}}{\sqrt{2} M_{pl}} $, which is numerically satisfied, both being of the order of $4\times 10^{-8} M_{pl}$.

And, using that at the reheating time (i.e., when the energy density of the scalar field and the one of the relativistic plasma coincide) 
the Hubble rate is given by $H_{rh}^2=\frac{2\rho_{rh}}{3M_{pl}^2}$, one gets 
\begin{eqnarray}
\varphi_{rh}=\varphi_{kin}+\sqrt{\frac{2}{3}}M_{pl}\ln\left( \frac{ H_{kin}}{\sqrt{\frac{\pi^2g_{rh}}{45}} \frac{T_{rh}^2}{M_{pl}}}\right)
\end{eqnarray} 
and
\begin{equation}
\dot{\varphi}_{rh}=\sqrt{\frac{\pi^2g_{rh}}{15}} T_{rh}^2,
\end{equation}
where we have used that  the energy density and the temperature are related via  the formula $\rho_{rh}=\frac{\pi^2}{30}g_{rh}T_{rh}^4$, where the number of degrees of freedom for the Standard Model is $g_{rh}=106.75$ \cite{Rehagen:2015zma}. Because of the smoothness of the potential,
since the gravitational particle production 
\cite{Ford:1986sy,Haro:2018zdb,Hashiba:2018iff, Chung:1998zb,Chung:2001cb}
only works for potentials with an abrupt phase transition leading  to a non-adiabatic process which allows the production of particles (see for instance the Peebles-Vilenkin potential \cite{Peebles:1998qn}), 
we consider ``Instant Preheating'' \cite{Felder:1998vq,Felder:1999pv,Haro:2018jtb, Dimopoulos:2017tud} and, thus, we will choose  as the reheating temperature $T_{rh}\cong  10^{9}$ GeV, which is its usual value when the mechanism to reheat the universe is this one.

Effectively, considering a massless scalar $X$-field conformally coupled with gravity and interacting with the inflaton field as follows, ${\mathcal L}_{int}=-\frac{1}{2}g\varphi^2X^2$ \cite{Felder:1998vq,Felder:1999pv},
where $g$ is the dimensionless coupling constant and where the Enhanced Symmetry Point (ESP)
 has been chosen at $\varphi=0$ because, as we have already shown numerically, the beginning of the kination starts at $\varphi_{kin}\sim -0.03M_{pl}$. Then, at the beginning of the kination, the adiabaticity is broken and $X$-particles are produced with a number density  equal to \cite{Haro:2019ndy}
\begin{eqnarray}
n_{X,kin}=\frac{g^{3/2}\dot{\varphi}_{kin}^{3/2}}{8\pi^3}
\end{eqnarray}
and, since these particles acquire a very heavy  effective mass equal to $gM_{pl}$, in order to reheat the universe they have to decay into lighter ones forming a relativistic plasma, whose energy density will eventually dominate the one of the inflaton field (recall that during kination the energy density of the field decays as $a^{-6}$ and the one of the relativistic plasma as $a^{-4}$), obtaining a reheated universe with a  
reheating temperature  given by \cite{Haro:2018jtb}
\begin{eqnarray}
T_{rh}=\left( \frac{30}{g_*\pi^2} \right)^{1/4}\rho_{X,dec}^{1/4}\sqrt{\frac{\rho_{X,dec}}{\rho_{\varphi, dec}}}\\
\sim 10^{14} g^{15/8}\left(\frac{M_{pl}}{\Gamma} \right)^{1/4} \mbox{ GeV},\nonumber
\end{eqnarray}
where $g_*=106.75$ are the degrees of freedom for the Standard Model, $\Gamma$ is the decay rate and the sub-index ``dec'' denotes the moment when the $X$-field decays completely.

Assuming now that the $X$-field decays into fermions via a Yukawa type of interaction
$h\psi\bar{\psi}X$ with decay rate $\Gamma=\frac{h^2gM_{pl}}{8\pi}$, where $h$ is a dimensionless coupling constant, one gets
\begin{eqnarray}
T_{rh}\sim 10^{14}g^{13/8}h^{-1/2}\mbox{ GeV},
\end{eqnarray}
which leads for the narrow range of viable parameters $g$ and $h$ \cite{Haro:2018jtb} to a reheating temperature around $10^9$ GeV.


So, finally, at the beginning of the radiation era we have
\begin{eqnarray}
 \varphi_{rh}\cong 20 M_{pl} \qquad {\dot{\varphi}_{rh}\cong 1.4\times 10^{-18}} M_{pl}^2.
\end{eqnarray}

\section{Numerical simulation}
\label{sec:numerics}

First of all, we consider the central values obtained in \cite{Ade:2015xua} (see the second column in Table $4$ of \cite{Ade:2015xua}) of  the red-shift at the matter-radiation equality $z_{eq}=3365$,
the present value of the ratio of the matter energy density to the critical one $\Omega_{m,0}=0.308$, and, once again,  $H_0=67.81\; \mbox{Km/sec/Mpc}=5.94\times 10^{-61} M_{pl}$.
Then, the present value of the matter energy density is $\rho_{m,0}=3H_0^2M_{pl}^2\Omega_{m,0}=3.26\times 10^{-121} M_{pl}^4$, and at matter-radiation equality we will 
have $\rho_{eq}=2\rho_{m,0}(1+z_{eq})^3=2.48\times 10^{-110} M_{pl}^4=8.8\times 10^{-1} \mbox{eV}^4$. 
So,
at the beginning of matter-radiation equality the energy density of the matter  and radiation will be
 $\rho_{m,eq}=\rho_{r,eq}=\rho_{eq}/2\cong 4.4\times 10^{-1} \mbox{eV}^4$.

In this way,  
the dynamical equations after the beginning of the radiation can be easily obtained using as a time variable
$N\equiv -\ln(1+z)=\ln\left( \frac{a}{a_0}\right)$. Recasting the  energy density of radiation and matter respectively as functions of $N$, we get
\begin{eqnarray}
\hspace{-0.5cm}\rho_{m}(a)={\rho_{m,eq}}\left(\frac{a_{eq}}{a}  \right)^3\rightarrow \rho_{m}(N)={\rho_{m,eq}}e^{3(N_{eq}-N)}
 \end{eqnarray}
and
\begin{eqnarray}
\hspace{-0.5cm}\rho_{r}(a)={\rho_{r,eq}}\left(\frac{a_{eq}}{a}  \right)^4\rightarrow \rho_{r}(N)= {\rho_{r,eq}}e^{4(N_{eq}-N)},
\end{eqnarray}
where  
$N_{eq}\cong -8.121$ denotes the value of the time $N$ at the beginning of the matter-radiation equality. To obtain the dynamical system for this scalar field model, we will introduce the dimensionless variables
 \begin{equation}
 x=\frac{\varphi}{M_{pl}},  \quad  y=\frac{\dot{\varphi}}{H_0 M_{pl}}.     
 \end{equation}
Taking into account the conservation equation $\ddot{\varphi}+3H\dot{\varphi}+V_{\varphi}=0$, one arrives at the following dynamical system,
 \begin{equation}
 \label{system}
     x' = y/\bar H, \quad y' = -3y-\bar{V}_x/\bar{H},
 \end{equation}
 where the prime is the derivative with respect to $N$, $\bar{H}=\frac{H}{H_0}$   and $\bar{V}=\frac{V}{H_0^2M_{pl}^2}$.  It is not difficult to see that  one can write  
 \begin{eqnarray}
 \bar{H}=\frac{1}{\sqrt{3}}\sqrt{ \frac{y^2}{2}+\bar{V}(x)+ \bar{\rho}_{r}(N)+\bar{\rho}_{m}(N) }~,
 \end{eqnarray}
where we have defined the dimensionless energy densities as
\begin{equation}
\bar{\rho}_{r}=\frac{\rho_{r}}{H_0^2M_{pl}^2}, \quad \bar{\rho}_{m}=\frac{\rho_{m}}{H_0^2M_{pl}^2}.     
\end{equation}
Finally,  we have to integrate the dynamical system (\ref{system}), with initial conditions $x(N_{rh})=x_{rh}= 20$ and $y(N_{rh})=y_{rh}= 2.4\times 10^{42}$
 imposing that $\bar{H}(0)=1$, where $N_{rh}$ denotes the beginning of reheating, which is obtained imposing that
\begin{eqnarray}
{\rho_{r,eq}}e^{4(N_{eq}-N_{rh})}= \frac{\pi^2}{30}g_{rh} T^4_{rh},\end{eqnarray}
that is,
\begin{eqnarray}
N_{rh}=N_{eq}-\frac{1}{4}\ln\left(\frac{g_{rh}}{g_{eq}}\right)-\ln\left(\frac{T_{rh}}{T_{eq}}\right)\cong -50.68,
\end{eqnarray} 
where we have used that $\rho_{eq,r}=\frac{\pi^2}{30}g_{eq} T^4_{eq}$ with $g_{eq}=3.36$ and, thus, $T_{eq}\cong 7.81\times 10^{-10}$ GeV.

\begin{figure*}[t!]
\begin{center}
\includegraphics[width=0.9\textwidth]{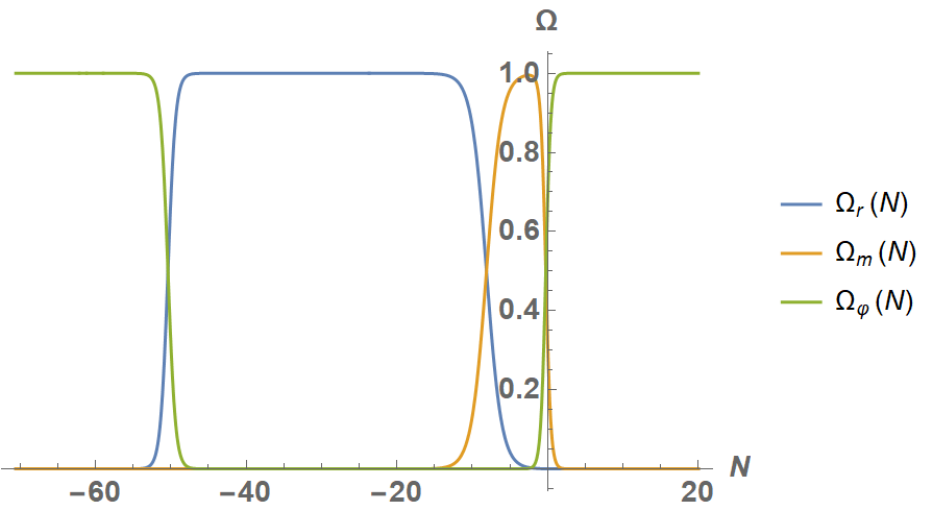}
\\
\includegraphics[width=0.9\textwidth]{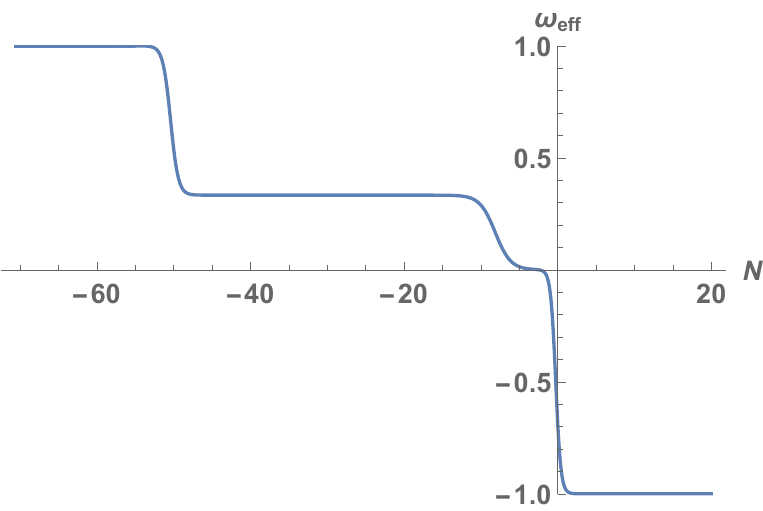}
\caption{\it{Upper: The density parameters $\Omega_m=\frac{\rho_m}{3H^2M_{pl}^2}$ (orange curve), $\Omega_r=\frac{\rho_r}{3H^2M_{pl}^2}$ (blue curve) and $\Omega_{\varphi}=\frac{\rho_{\varphi}}{3H^2M_{pl}^2}$, from kination to future times. Lower: The effective Equation of State parameter $w_{eff}$, from kination to future times. As one can see in the picture, after kination the universe enters in a large period of time where radiation dominates. Then, after the matter-radiation equality, the universe becomes matter-dominated and, finally, near the present, it enters in a new accelerated phase where $w_{eff}$ approaches $-1$.}} \label{fig:Omega}
\end{center}
\end{figure*}

We have numerically checked that, to obtain the condition $\bar{H}(0)=1$, the parameters $\gamma$ and $\xi$ have to be equal to $121.8$. Once these parameters have been properly selected, the obtained results are presented in Figure \ref{fig:Omega}.

\section{Other similar possible models}
\label{sec:othermodels}

In this section, we are going to test different models that resemble the one that has been considered so far.
An analogously built simplified model would be 
\begin{eqnarray}
V(\varphi)=\lambda M_{pl}^4 e^{-\alpha\arctan\left(\beta\frac{\varphi}{M_{pl}} \right)},
\end{eqnarray}
for $\alpha$ and $\beta$ being its positive parameters, where we have suppressed the $sinh$ function. Its slow-roll parameters $\epsilon$ and $\eta$ are
\begin{eqnarray}
\epsilon=\frac{1}{2}\left(\frac{\alpha\beta}{1+\left(\beta\frac{\varphi}{M_{pl}} \right)^2} \right)^2, \\ 
\eta=\left(\frac{\alpha\beta}{1+\left(\beta\frac{\varphi}{M_{pl}} \right)^2} \right)^2\left(1 + 2\frac{\beta}{\alpha}\frac{\varphi}{M_{pl}} \right)\nonumber.
\end{eqnarray}
Hence, the slow-roll parameter $\epsilon$ is also related to a Lorentzian distribution, in this case in function of $\varphi$ instead of $N$ and with an overall square involved, which makes this model an interesting case worth to study. Using that $\left|\beta\frac{\varphi_*}{M_{pl}}\right|\gg\max(1,|\alpha)$, we get that
\begin{eqnarray}
n_s\cong 1+4\frac{\alpha}{\beta}\left(\frac{M_{pl}}{\varphi_*} \right)^3, \quad
r\cong 8\left(\frac{\alpha}{\beta} \right)^2\left(\frac{M_{pl}}{\varphi_*}\right)^4.
\end{eqnarray}
Using the same approximations, one can find that 
\begin{eqnarray}
N=\frac{1}{M_{pl}}\int_{\varphi_*}^{\varphi_{END}}\frac{1}{\sqrt{2\epsilon}}d\varphi\cong -\frac{\beta}{3\alpha}\left(\frac{\varphi_*}{M_{pl}} \right)^3
\end{eqnarray}
and, therefore,
\begin{eqnarray}
n_s\cong 1-\frac{4}{3N} \qquad \text{and} \qquad r\cong 8 \left(\frac{\alpha}{9\beta N^2} \right)^{2/3}.
\end{eqnarray}
In order to study the viability of this model, let's start by fixing the value of the parameters in analogy to the main model considered in this work, namely $\beta=\gamma$ and $\alpha=\frac{2\xi}{\pi}$. In this case, the relation between the number of e-folds and the reheating temperature yields
\begin{eqnarray}
N+\frac{2}{3}\ln N = 53.13 + \ln\left(\frac{a_{END}}{a_{kin}}\right)-\frac{1}{3}\ln\left(\frac{T_{rh}}{M_{pl}}\right),
\end{eqnarray}
which leads to $57.6\lesssim N\lesssim 66.7$, for which the spectral index clearly falls outside of the allowed range.
So, this model does not work in the same way as we have shown for the previous one. However, this does not rule out that for other values of the parameters $\alpha$ and $\beta$ viability could be proved as well. The same applies to the model named ``arctan inflation''  introduced in \cite{Wang:1997cw},
\begin{eqnarray}
V(\varphi)=\lambda M_{pl}^4\left(1- \alpha\arctan\left(\beta\frac{\varphi}{M_{pl}}\right)\right),
\end{eqnarray}
{  whose viability was proved in Section $4.19$ of \cite{Martin:2013tda} for some given parameters $\alpha$ and $\beta$, which does not contradict our statement.}

To finish this section a final comment is in order. One could also use the original model obtained in \cite{Benisty:2020xqm}, 
\begin{eqnarray}\label{exp1}
\hspace{-0.5cm}V(\varphi)=\lambda M_{pl}^4\exp\left[-\frac{2\xi}{\pi}\arctan\left(\sinh
\left(\gamma\varphi/M_{pl} \right)  \right)\right]\boldsymbol{\cdot} \nonumber\\
\left(1-\frac{2\gamma^2\xi^2}{3\pi^3}\frac{1}{\cosh
\left(\gamma\varphi/M_{pl} \right) } \right),
\end{eqnarray}
which is negative around $\varphi\cong 0$. However, it leads
 to the exact same results, given that the only change is the behavior of the potential for $\varphi\cong 0$.
 
 Effectively, when the pivot scale leaves the Hubble radius one has
 \begin{eqnarray}
 \left(1-\frac{2\gamma^2\xi^2}{3\pi^3}\frac{1}{\cosh
\left(\gamma\varphi_*/M_{pl} \right) } \right)\cong
\left( 1- \frac{r}{64\pi} \right)\cong 1,
 \end{eqnarray}
 because $r\cong \frac{8}{N^2\gamma^2}\ll 1$. Thus, the last term of the potential (\ref{exp1}) does not affect to the power spectrum of perturbations.  In the same way, one can easily check that at the end of inflation the potential is positive. Therefore,
 defining once again that kination starts when $w_{\varphi}\cong 1$, which occurs when $\varphi_{kin}=0.073 M_{pl}$ (corresponding now to the time when the potential becomes positive again), everything works as expected.

\section{Cosmological Probes}
\label{sec:Data}
\begin{figure*}[t!]
\begin{center}
\includegraphics[width=15cm]{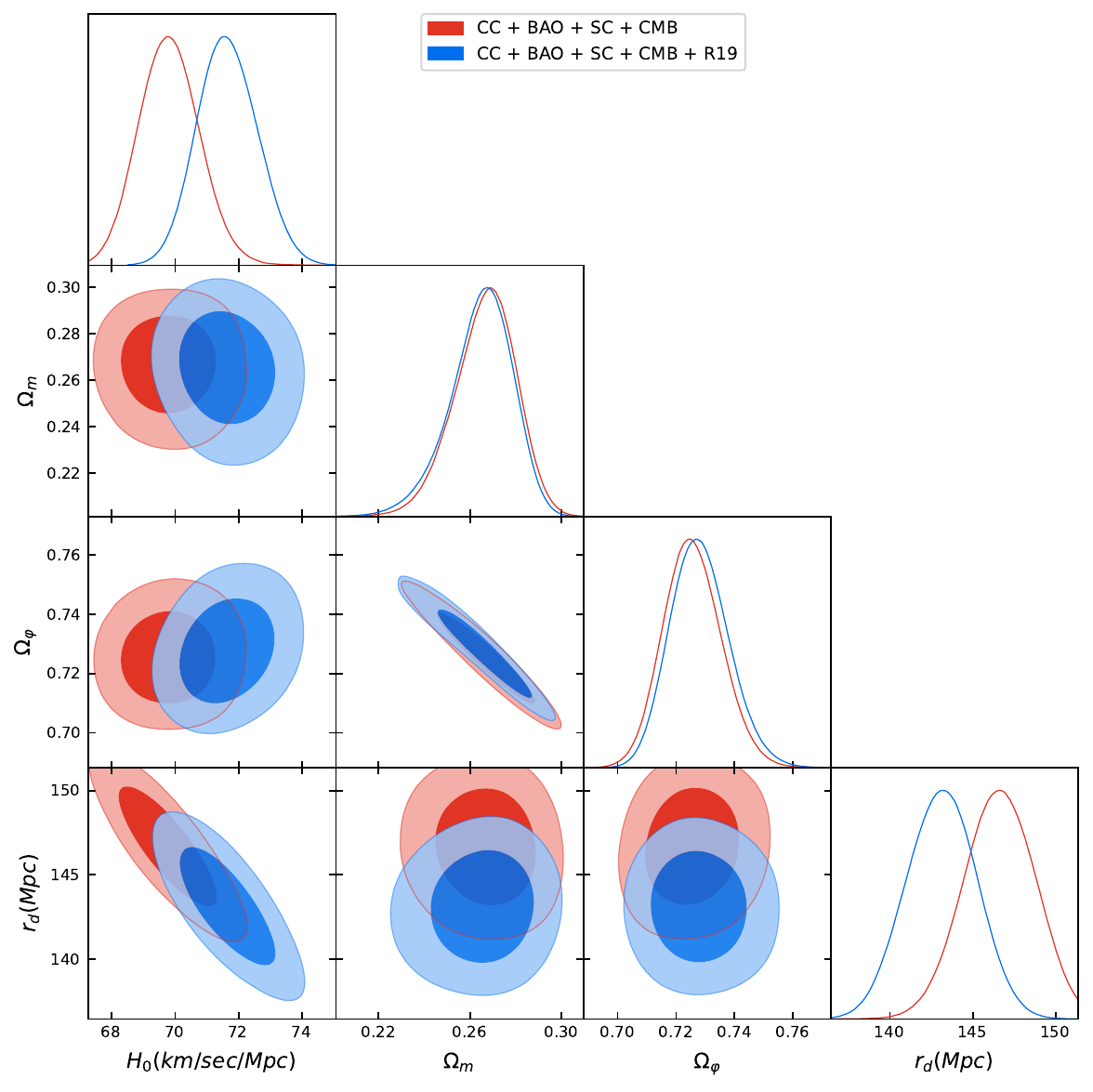}
\end{center}
\caption{\it{The posterior distribution for the LQI model with $1 \sigma$ and $2\sigma$. The data set include Baryon Acoustic Oscillations dataset, Cosmic Chronometers, the Hubble Diagram from Type Ia supernova, Quasars and Gamma Ray Bursts and the CMB. R19 denotes the Riess 2019 measurement of the Hubble constant as a Gaussian prior.}} \label{fig:fit}
\end{figure*}
\begin{table*}[t!]
\tabcolsep 5.5pt
\vspace{1mm}
\centering
\begin{tabular}{ccc} \hline \hline
Parameter & LQI & LQI + SH0ES\\ \hline \hline
$H_0 (km/sec/Mpc)$ & $70.06 \pm 1.123$ & $71.75 \pm 0.8885$ \\
$\varphi_0/M_{pl}$ & $22.72 \pm 1.541$ & $22.38 \pm 1.395$\\
$\dot{\varphi}_0/(H_0  M_{pl}) \, 10^{-71}  $ & $4.113 \pm 2.635$ & $5.279 \pm 2.675$ \\
$\Omega_m$ & $0.2679 \pm 0.01286$ &$0.2610 \pm 0.01647$ \\
{  $\Omega_{\varphi}$} & $0.7250 \pm 0.09131$ & $0.7304 \pm 0.01107$ \\
$\xi$ & $121.9 \pm 1.865$ & $122.0 \pm 1.94$ \\
$r_d (Mpc)$ &$145.8 \pm 2.363$& $143.0 \pm 1.957$\\
\hline\hline
\end{tabular}
\caption[]{\it{The best fit values for the discussed model for a uniform prior of the Hubble parameter and for a Gaussian prior that demonstrates the SH0ES measurement (Supernovae and H0 for the Dark Energy Equation of State). The values $\varphi_0$, $\dot{\varphi}_0$ denote the current values of the scalar field and its derivative.}}
\label{tab:Res}
\end{table*}

In order to constrain our model, we use a few data sets: 
\textbf{Cosmic Chronometers (CC)} exploit the evolution of differential ages of passive galaxies at different redshifts to directly constrain the Hubble
parameter \cite{Jimenez:2001gg}. We use uncorrelated 30 CC measurements of $H(z)$ discussed in \cite{Moresco:2012by,Moresco:2012jh,Moresco:2015cya,Moresco:2016mzx}. For \textbf{Standard Candles (SC)} we use measurements of the Pantheon Type Ia supernova dataset \cite{Scolnic:2017caz} that were collected in \cite{Anagnostopoulos:2020ctz} and the measurements from Quasars \cite{Roberts:2017nkm} and Gamma Ray Bursts \cite{Demianski:2016zxi}. The parameters of the models are to be fitted with by comparing the observed $\mu _{i}^{obs}$ value to the
theoretical $\mu _{i}^{th}$ value of the distance moduli, which is given by 
\begin{equation}
 \mu=m-M=5\log _{10}(D_{L})+\mu _{0},   
\end{equation}
where $m$ and $M$ are the apparent and absolute magnitudes and $\mu
_{0}=5\log \left( H_{0}^{-1}/Mpc\right) +25$ is the nuisance parameter that
has been marginalized. The distance moduli is given for different redshifts $\mu_i=\mu(z_i)$. The luminosity distance is defined by
\begin{eqnarray}
D_L(z) &=&\frac{c}{H_{0}}(1+z)\int_{0}^{z}\frac{dz^{\ast }}{%
E(z^{\ast })} ,
\end{eqnarray}%
where $E(z)=\frac{H(z)}{H_0}$. Here, we are assuming that $\Omega _{k}=0$ (flat space-time). 

We use uncorrelated data points from different \textbf{Baryon Acoustic Oscillations (BAO)} collected in \cite{Benisty:2020otr} from \cite{Percival:2009xn,Beutler:2011hx,Busca:2012bu,Anderson:2012sa,Seo:2012xy,Ross:2014qpa,Tojeiro:2014eea,Bautista:2017wwp,deCarvalho:2017xye,Ata:2017dya,Abbott:2017wcz,Molavi:2019mlh}. Studies of the BAO feature in the transverse direction provide a measurement of $D_H(z)/r_d = c/H(z)r_d$, where $r_d$ is the sound horizon at the drag epoch and it is taken as an independent parameter and with the comoving angular diameter distance \cite{Hogg:2020ktc,Martinelli:2020hud} being 
\begin{equation}
D_M= \int_0^z\frac{c \, dz'}{H(z')}.
\end{equation}
In our database we also use the angular diameter distance $D_A=D_M/(1+z)$ and $D_V(z)/r_d$, which is a combination of the BAO peak coordinates above, namely
\begin{equation}
    D_V(z) \equiv [ z D_H(z) D_M^2(z) ]^{1/3}.
\end{equation}
Finally we take the \textbf{CMB Distant Prior} measurements \cite{Chen:2018dbv}. The distance priors provide effective information of the CMB power spectrum in two aspects: the acoustic scale $l_A $
characterizes the CMB temperature power spectrum in the transverse direction, leading to the variation of the peak
spacing, and the “shift parameter” $R$ influences the CMB temperature spectrum along the line-of-sight direction,
affecting the heights of the peaks, which are defined as follows:
\begin{equation}
\begin{split}
l_A =  (1 + z) \frac{\pi D_A(z)}{r_s}, \quad R(z) = \frac{\sqrt{\Omega_m} H_0}{c}(1 + z)  D_A(z).  
\end{split}
\end{equation}
The observables that \cite{Chen:2018dbv} reports are:
\begin{equation}
R_{z} = 1.7502 \pm 0.0046,\quad l_{A} =  301.471\pm0.09, \quad n_s = 0.9649 \pm 0.0043 
\end{equation}
with a corresponding covariance matrix (see table I in \cite{Chen:2018dbv}). The points incorporate the expansion rate from the CMB epoch, and the observables from inflation. We also include other measurements from the late universe in addition to the CMB points. The combination yields a good test for the model with respect to the data.

In our analysis we used $r_s$ as independent parameter. We take the complete analyses that combine the likelihoods from all of the datasets. We use a nested sampler as it is implemented within the open-source packaged $Polychord$ \cite{Handley:2015fda} with the $GetDist$ package \cite{Lewis:2019xzd} to present the results. The prior we choose is with a uniform distribution, where $\Omega_{r} \in [0;1.]$, $\Omega_{m}\in[0.;1.]$, $\varphi_0\in [20;25]$, $\dot{\varphi}_0\in [0;10^{-70}]$  $\Omega_{\varphi}\in[0.;1.]$, $H_0\in [50;100]$Km/sec/Mpc, $\xi = \gamma = \in [100;130]$,  $r_s\in [130;160]$Mpc. The measurement of the Hubble constant yielding $H_0 = 74.03 \pm 1.42$ (km/s)/Mpc at $68\%$ CL by \cite{Riess:2019cxk} has been incorporated into our analysis as an additional prior (\textbf{R19}).

\begin{figure}[t!]
\begin{center}
\includegraphics[width=12cm]{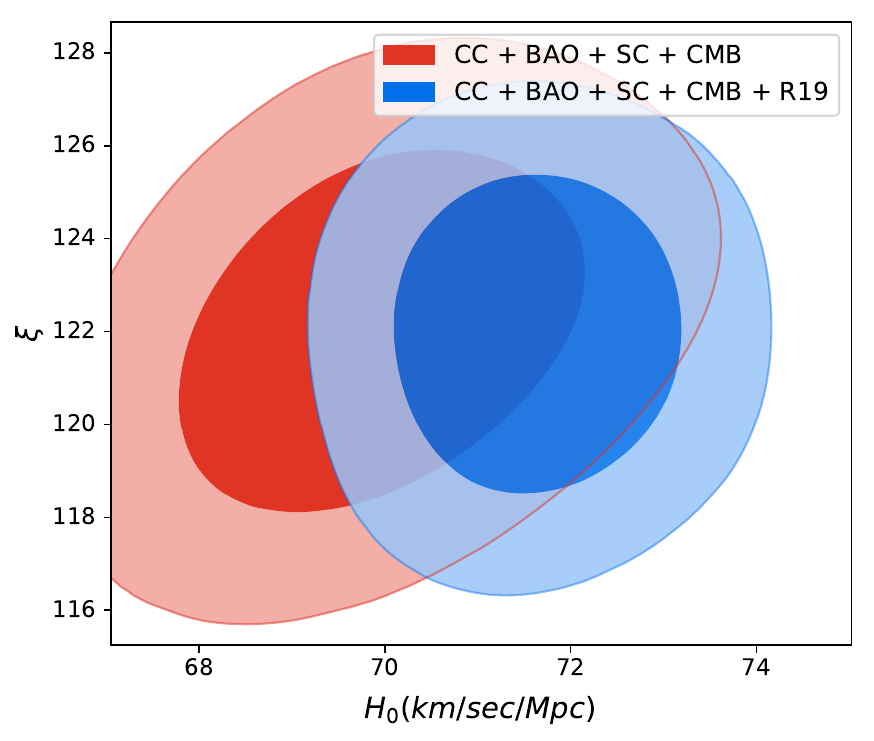}
\end{center}
\caption{\it{The posterior distribution for the LQI model with $1 \sigma$ and $2\sigma$, for the Hubble parameter vs. the parameter $\xi$. The data set include Baryon Acoustic Oscillations dataset, Cosmic Chronometers, the Hubble Diagram from Type Ia supernova, Quasars and Gamma Ray Bursts and the CMB. R19 denotes the Riess 2019 measurement of the Hubble constant as a Gaussian prior.}} \label{fig:fitH0}
\end{figure}

Figure \ref{fig:fit} shows the posterior distribution of the data fit with the best fit values at table \ref{tab:Res}. One can see that the Gaussian prior of the Hubble parameter does not change the results by much. For both cases the $\chi^2$ minimized value gives a good fit, since $\chi^2/Dof = [255.7/273,257.0/273] \sim 1$, where $Dof$ are the degrees of freedom for the $\chi^2$ distribution. The statement from the fit shows that the QI models that we discuss here are viable models and can describe early times as well as late times.



\section{Concluding remarks} \label{sec:summary}
In this paper we study the phenomenological implications of
a Lorentzian Quintessential Model, depending only on two parameters, where the reheating of the universe, due to the smoothness of the corresponding potential,  is produced  via the well-known {\it Instant Preheating} mechanism. We have  shown, analytically and numerically,  that for reasonable value of these parameters, this simple model is able to depict correctly our universe unifying its early and late time acceleration.  In fact, the model belongs to the class of the so-called $\alpha$-attractors and, thus, matches very well with the observational data  of the power spectrum of perturbation during inflation provided by the Planck's team.  It leads to a current dark energy density around  $70$\% of the total one.

In addition to the reheating constraints, we have tested the model with different measurements, some of them from the late universe such as Type Ia supernova, Gamma Ray Bursts and Quasars and the others from the Cosmic Microwave Background from the early universe. The model fits very well to the latest measurements and gives a reasonable scenario beyond the standard $\Lambda$CDM that includes the inflationary epoch. Further analysis of the $\alpha$ attractors with LQI background is studied in \cite{AresteSalo:2021wgb}.

\acknowledgments
We would like to thank Fotios Anagnostopoulos for helpful comments and advice. D. Benisty and E. I. Guendelman thank Ben Gurion University of the Negev for a great support. D. Benisty also thanks to the Grants Committee of the Rothschild and the Blavatnik Cambridge Fellowships for generous supports. The investigation { of J. de Haro} has been supported by MINECO (Spain) grant  MTM2017-84214-C2-1-P and  in part by the Catalan Government 2017-SGR-247. This work has been supported by the European COST actions CA15117 and CA18108.

\bibliographystyle{apsrev4-1}
\bibliography{ref}

\begin{thebibliography}{116}%
\makeatletter
\providecommand \@ifxundefined [1]{%
 \@ifx{#1\undefined}
}%
\providecommand \@ifnum [1]{%
 \ifnum #1\expandafter \@firstoftwo
 \else \expandafter \@secondoftwo
 \fi
}%
\providecommand \@ifx [1]{%
 \ifx #1\expandafter \@firstoftwo
 \else \expandafter \@secondoftwo
 \fi
}%
\providecommand \natexlab [1]{#1}%
\providecommand \enquote  [1]{``#1''}%
\providecommand \bibnamefont  [1]{#1}%
\providecommand \bibfnamefont [1]{#1}%
\providecommand \citenamefont [1]{#1}%
\providecommand \href@noop [0]{\@secondoftwo}%
\providecommand \href [0]{\begingroup \@sanitize@url \@href}%
\providecommand \@href[1]{\@@startlink{#1}\@@href}%
\providecommand \@@href[1]{\endgroup#1\@@endlink}%
\providecommand \@sanitize@url [0]{\catcode `\\12\catcode `\$12\catcode
  `\&12\catcode `\#12\catcode `\^12\catcode `\_12\catcode `\%12\relax}%
\providecommand \@@startlink[1]{}%
\providecommand \@@endlink[0]{}%
\providecommand \url  [0]{\begingroup\@sanitize@url \@url }%
\providecommand \@url [1]{\endgroup\@href {#1}{\urlprefix }}%
\providecommand \urlprefix  [0]{URL }%
\providecommand \Eprint [0]{\href }%
\providecommand \doibase [0]{http://dx.doi.org/}%
\providecommand \selectlanguage [0]{\@gobble}%
\providecommand \bibinfo  [0]{\@secondoftwo}%
\providecommand \bibfield  [0]{\@secondoftwo}%
\providecommand \translation [1]{[#1]}%
\providecommand \BibitemOpen [0]{}%
\providecommand \bibitemStop [0]{}%
\providecommand \bibitemNoStop [0]{.\EOS\space}%
\providecommand \EOS [0]{\spacefactor3000\relax}%
\providecommand \BibitemShut  [1]{\csname bibitem#1\endcsname}%
\let\auto@bib@innerbib\@empty
\bibitem [{\citenamefont {Benisty}\ and\ \citenamefont
  {Guendelman}(2020{\natexlab{a}})}]{Benisty:2020xqm}%
  \BibitemOpen
  \bibfield  {author} {\bibinfo {author} {\bibfnamefont {D.}~\bibnamefont
  {Benisty}}\ and\ \bibinfo {author} {\bibfnamefont {E.~I.}\ \bibnamefont
  {Guendelman}},\ }\href {\doibase 10.1142/S021827182042002X} {\bibfield
  {journal} {\bibinfo  {journal} {Int. J. Mod. Phys. D}\ }\textbf {\bibinfo
  {volume} {29}},\ \bibinfo {pages} {2042002} (\bibinfo {year}
  {2020}{\natexlab{a}})},\ \Eprint {http://arxiv.org/abs/2004.00339}
  {arXiv:2004.00339 [astro-ph.CO]} \BibitemShut {NoStop}%
\bibitem [{\citenamefont {Benisty}\ and\ \citenamefont
  {Guendelman}(2020{\natexlab{b}})}]{Benisty:2020qta}%
  \BibitemOpen
  \bibfield  {author} {\bibinfo {author} {\bibfnamefont {D.}~\bibnamefont
  {Benisty}}\ and\ \bibinfo {author} {\bibfnamefont {E.~I.}\ \bibnamefont
  {Guendelman}},\ }\href {\doibase 10.1140/epjc/s10052-020-8147-8} {\bibfield
  {journal} {\bibinfo  {journal} {Eur. Phys. J. C}\ }\textbf {\bibinfo {volume}
  {80}},\ \bibinfo {pages} {577} (\bibinfo {year} {2020}{\natexlab{b}})},\
  \Eprint {http://arxiv.org/abs/2006.04129} {arXiv:2006.04129 [astro-ph.CO]}
  \BibitemShut {NoStop}%
\bibitem [{\citenamefont {Weinberg}(1989)}]{Weinberg:1988cp}%
  \BibitemOpen
  \bibfield  {author} {\bibinfo {author} {\bibfnamefont {S.}~\bibnamefont
  {Weinberg}},\ }\href {\doibase 10.1103/RevModPhys.61.1} {\bibfield  {journal}
  {\bibinfo  {journal} {Rev. Mod. Phys.}\ }\textbf {\bibinfo {volume} {61}},\
  \bibinfo {pages} {1} (\bibinfo {year} {1989})},\ \bibinfo {note}
  {[,569(1988)]}\BibitemShut {NoStop}%
\bibitem [{\citenamefont {Lombriser}(2019)}]{Lombriser:2019jia}%
  \BibitemOpen
  \bibfield  {author} {\bibinfo {author} {\bibfnamefont {L.}~\bibnamefont
  {Lombriser}},\ }\href {\doibase 10.1016/j.physletb.2019.134804} {\bibfield
  {journal} {\bibinfo  {journal} {Phys. Lett.}\ }\textbf {\bibinfo {volume}
  {B797}},\ \bibinfo {pages} {134804} (\bibinfo {year} {2019})},\ \Eprint
  {http://arxiv.org/abs/1901.08588} {arXiv:1901.08588 [gr-qc]} \BibitemShut
  {NoStop}%
\bibitem [{\citenamefont {Frieman}\ \emph {et~al.}(2008)\citenamefont
  {Frieman}, \citenamefont {Turner},\ and\ \citenamefont
  {Huterer}}]{Frieman:2008sn}%
  \BibitemOpen
  \bibfield  {author} {\bibinfo {author} {\bibfnamefont {J.}~\bibnamefont
  {Frieman}}, \bibinfo {author} {\bibfnamefont {M.}~\bibnamefont {Turner}}, \
  and\ \bibinfo {author} {\bibfnamefont {D.}~\bibnamefont {Huterer}},\ }\href
  {\doibase 10.1146/annurev.astro.46.060407.145243} {\bibfield  {journal}
  {\bibinfo  {journal} {Ann. Rev. Astron. Astrophys.}\ }\textbf {\bibinfo
  {volume} {46}},\ \bibinfo {pages} {385} (\bibinfo {year} {2008})},\ \Eprint
  {http://arxiv.org/abs/0803.0982} {arXiv:0803.0982 [astro-ph]} \BibitemShut
  {NoStop}%
\bibitem [{\citenamefont {Riess}\ \emph {et~al.}(2019)\citenamefont {Riess},
  \citenamefont {Casertano}, \citenamefont {Yuan}, \citenamefont {Macri},\ and\
  \citenamefont {Scolnic}}]{Riess:2019cxk}%
  \BibitemOpen
  \bibfield  {author} {\bibinfo {author} {\bibfnamefont {A.~G.}\ \bibnamefont
  {Riess}}, \bibinfo {author} {\bibfnamefont {S.}~\bibnamefont {Casertano}},
  \bibinfo {author} {\bibfnamefont {W.}~\bibnamefont {Yuan}}, \bibinfo {author}
  {\bibfnamefont {L.~M.}\ \bibnamefont {Macri}}, \ and\ \bibinfo {author}
  {\bibfnamefont {D.}~\bibnamefont {Scolnic}},\ }\href {\doibase
  10.3847/1538-4357/ab1422} {\bibfield  {journal} {\bibinfo  {journal}
  {Astrophys. J.}\ }\textbf {\bibinfo {volume} {876}},\ \bibinfo {pages} {85}
  (\bibinfo {year} {2019})},\ \Eprint {http://arxiv.org/abs/1903.07603}
  {arXiv:1903.07603 [astro-ph.CO]} \BibitemShut {NoStop}%
\bibitem [{\citenamefont {Caldwell}\ \emph {et~al.}(1998)\citenamefont
  {Caldwell}, \citenamefont {Dave},\ and\ \citenamefont
  {Steinhardt}}]{Caldwell:1997ii}%
  \BibitemOpen
  \bibfield  {author} {\bibinfo {author} {\bibfnamefont {R.~R.}\ \bibnamefont
  {Caldwell}}, \bibinfo {author} {\bibfnamefont {R.}~\bibnamefont {Dave}}, \
  and\ \bibinfo {author} {\bibfnamefont {P.~J.}\ \bibnamefont {Steinhardt}},\
  }\href {\doibase 10.1103/PhysRevLett.80.1582} {\bibfield  {journal} {\bibinfo
   {journal} {Phys. Rev. Lett.}\ }\textbf {\bibinfo {volume} {80}},\ \bibinfo
  {pages} {1582} (\bibinfo {year} {1998})},\ \Eprint
  {http://arxiv.org/abs/astro-ph/9708069} {arXiv:astro-ph/9708069 [astro-ph]}
  \BibitemShut {NoStop}%
\bibitem [{\citenamefont {Ratra}\ and\ \citenamefont
  {Peebles}(1988)}]{Ratra:1987rm}%
  \BibitemOpen
  \bibfield  {author} {\bibinfo {author} {\bibfnamefont {B.}~\bibnamefont
  {Ratra}}\ and\ \bibinfo {author} {\bibfnamefont {P.~J.~E.}\ \bibnamefont
  {Peebles}},\ }\href {\doibase 10.1103/PhysRevD.37.3406} {\bibfield  {journal}
  {\bibinfo  {journal} {Phys. Rev.}\ }\textbf {\bibinfo {volume} {D37}},\
  \bibinfo {pages} {3406} (\bibinfo {year} {1988})}\BibitemShut {NoStop}%
\bibitem [{\citenamefont {Peebles}\ and\ \citenamefont
  {Ratra}(1988)}]{Peebles:1987ek}%
  \BibitemOpen
  \bibfield  {author} {\bibinfo {author} {\bibfnamefont {P.~J.~E.}\
  \bibnamefont {Peebles}}\ and\ \bibinfo {author} {\bibfnamefont
  {B.}~\bibnamefont {Ratra}},\ }\href {\doibase 10.1086/185100} {\bibfield
  {journal} {\bibinfo  {journal} {Astrophys. J. Lett.}\ }\textbf {\bibinfo
  {volume} {325}},\ \bibinfo {pages} {L17} (\bibinfo {year}
  {1988})}\BibitemShut {NoStop}%
\bibitem [{\citenamefont {Barreiro}\ \emph {et~al.}(2000)\citenamefont
  {Barreiro}, \citenamefont {Copeland},\ and\ \citenamefont
  {Nunes}}]{Barreiro:1999zs}%
  \BibitemOpen
  \bibfield  {author} {\bibinfo {author} {\bibfnamefont {T.}~\bibnamefont
  {Barreiro}}, \bibinfo {author} {\bibfnamefont {E.~J.}\ \bibnamefont
  {Copeland}}, \ and\ \bibinfo {author} {\bibfnamefont {N.~J.}\ \bibnamefont
  {Nunes}},\ }\href {\doibase 10.1103/PhysRevD.61.127301} {\bibfield  {journal}
  {\bibinfo  {journal} {Phys. Rev. D}\ }\textbf {\bibinfo {volume} {61}},\
  \bibinfo {pages} {127301} (\bibinfo {year} {2000})},\ \Eprint
  {http://arxiv.org/abs/astro-ph/9910214} {arXiv:astro-ph/9910214} \BibitemShut
  {NoStop}%
\bibitem [{\citenamefont {Carroll}(1998)}]{Carroll:1998zi}%
  \BibitemOpen
  \bibfield  {author} {\bibinfo {author} {\bibfnamefont {S.~M.}\ \bibnamefont
  {Carroll}},\ }\href {\doibase 10.1103/PhysRevLett.81.3067} {\bibfield
  {journal} {\bibinfo  {journal} {Phys. Rev. Lett.}\ }\textbf {\bibinfo
  {volume} {81}},\ \bibinfo {pages} {3067} (\bibinfo {year} {1998})},\ \Eprint
  {http://arxiv.org/abs/astro-ph/9806099} {arXiv:astro-ph/9806099 [astro-ph]}
  \BibitemShut {NoStop}%
\bibitem [{\citenamefont {Chiba}(1999)}]{Chiba:1999wt}%
  \BibitemOpen
  \bibfield  {author} {\bibinfo {author} {\bibfnamefont {T.}~\bibnamefont
  {Chiba}},\ }\href {\doibase 10.1103/PhysRevD.60.083508} {\bibfield  {journal}
  {\bibinfo  {journal} {Phys. Rev. D}\ }\textbf {\bibinfo {volume} {60}},\
  \bibinfo {pages} {083508} (\bibinfo {year} {1999})},\ \Eprint
  {http://arxiv.org/abs/gr-qc/9903094} {arXiv:gr-qc/9903094} \BibitemShut
  {NoStop}%
\bibitem [{\citenamefont {Sahni}\ and\ \citenamefont
  {Wang}(2000)}]{Sahni:1999qe}%
  \BibitemOpen
  \bibfield  {author} {\bibinfo {author} {\bibfnamefont {V.}~\bibnamefont
  {Sahni}}\ and\ \bibinfo {author} {\bibfnamefont {L.-M.}\ \bibnamefont
  {Wang}},\ }\href {\doibase 10.1103/PhysRevD.62.103517} {\bibfield  {journal}
  {\bibinfo  {journal} {Phys. Rev.}\ }\textbf {\bibinfo {volume} {D62}},\
  \bibinfo {pages} {103517} (\bibinfo {year} {2000})},\ \Eprint
  {http://arxiv.org/abs/astro-ph/9910097} {arXiv:astro-ph/9910097 [astro-ph]}
  \BibitemShut {NoStop}%
\bibitem [{\citenamefont {Krishnan}\ \emph
  {et~al.}(2021{\natexlab{a}})\citenamefont {Krishnan}, \citenamefont
  {Colgain}, \citenamefont {Sheikh-Jabbari},\ and\ \citenamefont
  {Yang}}]{Krishnan:2020vaf}%
  \BibitemOpen
  \bibfield  {author} {\bibinfo {author} {\bibfnamefont {C.}~\bibnamefont
  {Krishnan}}, \bibinfo {author} {\bibfnamefont {E.~O.}\ \bibnamefont
  {Colgain}}, \bibinfo {author} {\bibfnamefont {M.~M.}\ \bibnamefont
  {Sheikh-Jabbari}}, \ and\ \bibinfo {author} {\bibfnamefont {T.}~\bibnamefont
  {Yang}},\ }\href {\doibase 10.1103/PhysRevD.103.103509} {\bibfield  {journal}
  {\bibinfo  {journal} {Phys. Rev.}\ }\textbf {\bibinfo {volume} {D103}},\
  \bibinfo {pages} {103509} (\bibinfo {year} {2021}{\natexlab{a}})},\ \Eprint
  {http://arxiv.org/abs/2011.02858} {arXiv:2011.02858 [astro-ph.CO]}
  \BibitemShut {NoStop}%
\bibitem [{\citenamefont {Krishnan}\ \emph
  {et~al.}(2021{\natexlab{b}})\citenamefont {Krishnan}, \citenamefont
  {Mohayaee}, \citenamefont {Colgáin}, \citenamefont {Sheikh-Jabbari},\ and\
  \citenamefont {Yin}}]{Krishnan:2021dyb}%
  \BibitemOpen
  \bibfield  {author} {\bibinfo {author} {\bibfnamefont {C.}~\bibnamefont
  {Krishnan}}, \bibinfo {author} {\bibfnamefont {R.}~\bibnamefont {Mohayaee}},
  \bibinfo {author} {\bibfnamefont {E.~O.}\ \bibnamefont {Colgáin}}, \bibinfo
  {author} {\bibfnamefont {M.~M.}\ \bibnamefont {Sheikh-Jabbari}}, \ and\
  \bibinfo {author} {\bibfnamefont {L.}~\bibnamefont {Yin}},\ }\href@noop {} {\
   (\bibinfo {year} {2021}{\natexlab{b}})},\ \Eprint
  {http://arxiv.org/abs/2105.09790} {arXiv:2105.09790 [astro-ph.CO]}
  \BibitemShut {NoStop}%
\bibitem [{\citenamefont {Guth}(1981)}]{Guth:1980zm}%
  \BibitemOpen
  \bibfield  {author} {\bibinfo {author} {\bibfnamefont {A.~H.}\ \bibnamefont
  {Guth}},\ }\href {\doibase 10.1103/PhysRevD.23.347} {\bibfield  {journal}
  {\bibinfo  {journal} {Phys. Rev.}\ }\textbf {\bibinfo {volume} {D23}},\
  \bibinfo {pages} {347} (\bibinfo {year} {1981})},\ \bibinfo {note} {[Adv.
  Ser. Astrophys. Cosmol.3,139(1987)]}\BibitemShut {NoStop}%
\bibitem [{\citenamefont {Linde}(1982)}]{Linde:1981mu}%
  \BibitemOpen
  \bibfield  {author} {\bibinfo {author} {\bibfnamefont {A.~D.}\ \bibnamefont
  {Linde}},\ }\bibfield  {booktitle} {\emph {\bibinfo {booktitle} {{QUANTUM
  COSMOLOGY}}},\ }\href {\doibase 10.1016/0370-2693(82)91219-9} {\bibfield
  {journal} {\bibinfo  {journal} {Phys. Lett.}\ }\textbf {\bibinfo {volume}
  {108B}},\ \bibinfo {pages} {389} (\bibinfo {year} {1982})},\ \bibinfo {note}
  {[Adv. Ser. Astrophys. Cosmol.3,149(1987)]}\BibitemShut {NoStop}%
\bibitem [{\citenamefont {Starobinsky}(1980)}]{Starobinsky:1980te}%
  \BibitemOpen
  \bibfield  {author} {\bibinfo {author} {\bibfnamefont {A.~A.}\ \bibnamefont
  {Starobinsky}},\ }\href {\doibase 10.1016/0370-2693(80)90670-X} {\bibfield
  {journal} {\bibinfo  {journal} {Phys. Lett.}\ }\textbf {\bibinfo {volume}
  {91B}},\ \bibinfo {pages} {99} (\bibinfo {year} {1980})},\ \bibinfo {note}
  {771 (1980)}\BibitemShut {NoStop}%
\bibitem [{\citenamefont {Peebles}\ and\ \citenamefont
  {Vilenkin}(1999)}]{Peebles:1998qn}%
  \BibitemOpen
  \bibfield  {author} {\bibinfo {author} {\bibfnamefont {P.~J.~E.}\
  \bibnamefont {Peebles}}\ and\ \bibinfo {author} {\bibfnamefont
  {A.}~\bibnamefont {Vilenkin}},\ }\href {\doibase 10.1103/PhysRevD.59.063505}
  {\bibfield  {journal} {\bibinfo  {journal} {Phys. Rev.}\ }\textbf {\bibinfo
  {volume} {D59}},\ \bibinfo {pages} {063505} (\bibinfo {year} {1999})},\
  \Eprint {http://arxiv.org/abs/astro-ph/9810509} {arXiv:astro-ph/9810509
  [astro-ph]} \BibitemShut {NoStop}%
\bibitem [{\citenamefont {Dimopoulos}\ and\ \citenamefont
  {Valle}(2002)}]{Dimopoulos:2001ix}%
  \BibitemOpen
  \bibfield  {author} {\bibinfo {author} {\bibfnamefont {K.}~\bibnamefont
  {Dimopoulos}}\ and\ \bibinfo {author} {\bibfnamefont {J.~W.~F.}\ \bibnamefont
  {Valle}},\ }\href {\doibase 10.1016/S0927-6505(02)00115-9} {\bibfield
  {journal} {\bibinfo  {journal} {Astropart. Phys.}\ }\textbf {\bibinfo
  {volume} {18}},\ \bibinfo {pages} {287} (\bibinfo {year} {2002})},\ \Eprint
  {http://arxiv.org/abs/astro-ph/0111417} {arXiv:astro-ph/0111417 [astro-ph]}
  \BibitemShut {NoStop}%
\bibitem [{\citenamefont {Dimopoulos}\ and\ \citenamefont
  {Owen}(2017)}]{Dimopoulos:2017zvq}%
  \BibitemOpen
  \bibfield  {author} {\bibinfo {author} {\bibfnamefont {K.}~\bibnamefont
  {Dimopoulos}}\ and\ \bibinfo {author} {\bibfnamefont {C.}~\bibnamefont
  {Owen}},\ }\href {\doibase 10.1088/1475-7516/2017/06/027} {\bibfield
  {journal} {\bibinfo  {journal} {JCAP}\ }\textbf {\bibinfo {volume} {1706}},\
  \bibinfo {pages} {027} (\bibinfo {year} {2017})},\ \Eprint
  {http://arxiv.org/abs/1703.00305} {arXiv:1703.00305 [gr-qc]} \BibitemShut
  {NoStop}%
\bibitem [{\citenamefont {Hossain}\ \emph
  {et~al.}(2014{\natexlab{a}})\citenamefont {Hossain}, \citenamefont
  {Myrzakulov}, \citenamefont {Sami},\ and\ \citenamefont
  {Saridakis}}]{Hossain:2014coa}%
  \BibitemOpen
  \bibfield  {author} {\bibinfo {author} {\bibfnamefont {M.~W.}\ \bibnamefont
  {Hossain}}, \bibinfo {author} {\bibfnamefont {R.}~\bibnamefont {Myrzakulov}},
  \bibinfo {author} {\bibfnamefont {M.}~\bibnamefont {Sami}}, \ and\ \bibinfo
  {author} {\bibfnamefont {E.~N.}\ \bibnamefont {Saridakis}},\ }\href {\doibase
  10.1103/PhysRevD.89.123513} {\bibfield  {journal} {\bibinfo  {journal} {Phys.
  Rev.}\ }\textbf {\bibinfo {volume} {D89}},\ \bibinfo {pages} {123513}
  (\bibinfo {year} {2014}{\natexlab{a}})},\ \Eprint
  {http://arxiv.org/abs/1404.1445} {arXiv:1404.1445 [gr-qc]} \BibitemShut
  {NoStop}%
\bibitem [{\citenamefont {Hossain}\ \emph
  {et~al.}(2014{\natexlab{b}})\citenamefont {Hossain}, \citenamefont
  {Myrzakulov}, \citenamefont {Sami},\ and\ \citenamefont
  {Saridakis}}]{Hossain:2014xha}%
  \BibitemOpen
  \bibfield  {author} {\bibinfo {author} {\bibfnamefont {M.~W.}\ \bibnamefont
  {Hossain}}, \bibinfo {author} {\bibfnamefont {R.}~\bibnamefont {Myrzakulov}},
  \bibinfo {author} {\bibfnamefont {M.}~\bibnamefont {Sami}}, \ and\ \bibinfo
  {author} {\bibfnamefont {E.~N.}\ \bibnamefont {Saridakis}},\ }\href {\doibase
  10.1103/PhysRevD.90.023512} {\bibfield  {journal} {\bibinfo  {journal} {Phys.
  Rev. D}\ }\textbf {\bibinfo {volume} {90}},\ \bibinfo {pages} {023512}
  (\bibinfo {year} {2014}{\natexlab{b}})},\ \Eprint
  {http://arxiv.org/abs/1402.6661} {arXiv:1402.6661 [gr-qc]} \BibitemShut
  {NoStop}%
\bibitem [{\citenamefont {Wali~Hossain}\ \emph {et~al.}(2015)\citenamefont
  {Wali~Hossain}, \citenamefont {Myrzakulov}, \citenamefont {Sami},\ and\
  \citenamefont {Saridakis}}]{Hossain:2014zma}%
  \BibitemOpen
  \bibfield  {author} {\bibinfo {author} {\bibfnamefont {M.}~\bibnamefont
  {Wali~Hossain}}, \bibinfo {author} {\bibfnamefont {R.}~\bibnamefont
  {Myrzakulov}}, \bibinfo {author} {\bibfnamefont {M.}~\bibnamefont {Sami}}, \
  and\ \bibinfo {author} {\bibfnamefont {E.~N.}\ \bibnamefont {Saridakis}},\
  }\href {\doibase 10.1142/S0218271815300141} {\bibfield  {journal} {\bibinfo
  {journal} {Int. J. Mod. Phys. D}\ }\textbf {\bibinfo {volume} {24}},\
  \bibinfo {pages} {1530014} (\bibinfo {year} {2015})},\ \Eprint
  {http://arxiv.org/abs/1410.6100} {arXiv:1410.6100 [gr-qc]} \BibitemShut
  {NoStop}%
\bibitem [{\citenamefont {Haro}\ \emph
  {et~al.}(2019{\natexlab{a}})\citenamefont {Haro}, \citenamefont {Amorós},\
  and\ \citenamefont {Pan}}]{Haro:2019gsv}%
  \BibitemOpen
  \bibfield  {author} {\bibinfo {author} {\bibfnamefont {J.}~\bibnamefont
  {Haro}}, \bibinfo {author} {\bibfnamefont {J.}~\bibnamefont {Amorós}}, \
  and\ \bibinfo {author} {\bibfnamefont {S.}~\bibnamefont {Pan}},\ }\href
  {\doibase 10.1140/epjc/s10052-019-7012-0} {\bibfield  {journal} {\bibinfo
  {journal} {Eur. Phys. J.}\ }\textbf {\bibinfo {volume} {C79}},\ \bibinfo
  {pages} {505} (\bibinfo {year} {2019}{\natexlab{a}})},\ \Eprint
  {http://arxiv.org/abs/1901.00167} {arXiv:1901.00167 [gr-qc]} \BibitemShut
  {NoStop}%
\bibitem [{\citenamefont {de~Haro}\ \emph
  {et~al.}(2016{\natexlab{a}})\citenamefont {de~Haro}, \citenamefont
  {Amorós},\ and\ \citenamefont {Pan}}]{deHaro:2016hpl}%
  \BibitemOpen
  \bibfield  {author} {\bibinfo {author} {\bibfnamefont {J.}~\bibnamefont
  {de~Haro}}, \bibinfo {author} {\bibfnamefont {J.}~\bibnamefont {Amorós}}, \
  and\ \bibinfo {author} {\bibfnamefont {S.}~\bibnamefont {Pan}},\ }\href
  {\doibase 10.1103/PhysRevD.93.084018} {\bibfield  {journal} {\bibinfo
  {journal} {Phys. Rev.}\ }\textbf {\bibinfo {volume} {D93}},\ \bibinfo {pages}
  {084018} (\bibinfo {year} {2016}{\natexlab{a}})},\ \Eprint
  {http://arxiv.org/abs/1601.08175} {arXiv:1601.08175 [gr-qc]} \BibitemShut
  {NoStop}%
\bibitem [{\citenamefont {Geng}\ \emph {et~al.}(2015)\citenamefont {Geng},
  \citenamefont {Hossain}, \citenamefont {Myrzakulov}, \citenamefont {Sami},\
  and\ \citenamefont {Saridakis}}]{Geng:2015fla}%
  \BibitemOpen
  \bibfield  {author} {\bibinfo {author} {\bibfnamefont {C.-Q.}\ \bibnamefont
  {Geng}}, \bibinfo {author} {\bibfnamefont {M.~W.}\ \bibnamefont {Hossain}},
  \bibinfo {author} {\bibfnamefont {R.}~\bibnamefont {Myrzakulov}}, \bibinfo
  {author} {\bibfnamefont {M.}~\bibnamefont {Sami}}, \ and\ \bibinfo {author}
  {\bibfnamefont {E.~N.}\ \bibnamefont {Saridakis}},\ }\href {\doibase
  10.1103/PhysRevD.92.023522} {\bibfield  {journal} {\bibinfo  {journal} {Phys.
  Rev.}\ }\textbf {\bibinfo {volume} {D92}},\ \bibinfo {pages} {023522}
  (\bibinfo {year} {2015})},\ \Eprint {http://arxiv.org/abs/1502.03597}
  {arXiv:1502.03597 [gr-qc]} \BibitemShut {NoStop}%
\bibitem [{\citenamefont {Geng}\ \emph {et~al.}(2017)\citenamefont {Geng},
  \citenamefont {Lee}, \citenamefont {Sami}, \citenamefont {Saridakis},\ and\
  \citenamefont {Starobinsky}}]{Geng:2017mic}%
  \BibitemOpen
  \bibfield  {author} {\bibinfo {author} {\bibfnamefont {C.-Q.}\ \bibnamefont
  {Geng}}, \bibinfo {author} {\bibfnamefont {C.-C.}\ \bibnamefont {Lee}},
  \bibinfo {author} {\bibfnamefont {M.}~\bibnamefont {Sami}}, \bibinfo {author}
  {\bibfnamefont {E.~N.}\ \bibnamefont {Saridakis}}, \ and\ \bibinfo {author}
  {\bibfnamefont {A.~A.}\ \bibnamefont {Starobinsky}},\ }\href {\doibase
  10.1088/1475-7516/2017/06/011} {\bibfield  {journal} {\bibinfo  {journal}
  {JCAP}\ }\textbf {\bibinfo {volume} {1706}},\ \bibinfo {pages} {011}
  (\bibinfo {year} {2017})},\ \Eprint {http://arxiv.org/abs/1705.01329}
  {arXiv:1705.01329 [gr-qc]} \BibitemShut {NoStop}%
\bibitem [{\citenamefont {de~Haro}\ and\ \citenamefont
  {Elizalde}(2016)}]{deHaro:2016hsh}%
  \BibitemOpen
  \bibfield  {author} {\bibinfo {author} {\bibfnamefont {J.}~\bibnamefont
  {de~Haro}}\ and\ \bibinfo {author} {\bibfnamefont {E.}~\bibnamefont
  {Elizalde}},\ }\href {\doibase 10.1007/s10714-016-2072-z} {\bibfield
  {journal} {\bibinfo  {journal} {Gen. Rel. Grav.}\ }\textbf {\bibinfo {volume}
  {48}},\ \bibinfo {pages} {77} (\bibinfo {year} {2016})},\ \Eprint
  {http://arxiv.org/abs/1602.03433} {arXiv:1602.03433 [gr-qc]} \BibitemShut
  {NoStop}%
\bibitem [{\citenamefont {de~Haro}(2017)}]{deHaro:2016ftq}%
  \BibitemOpen
  \bibfield  {author} {\bibinfo {author} {\bibfnamefont {J.}~\bibnamefont
  {de~Haro}},\ }\href {\doibase 10.1007/s10714-016-2173-8} {\bibfield
  {journal} {\bibinfo  {journal} {Gen. Rel. Grav.}\ }\textbf {\bibinfo {volume}
  {49}},\ \bibinfo {pages} {6} (\bibinfo {year} {2017})},\ \Eprint
  {http://arxiv.org/abs/1602.07138} {arXiv:1602.07138 [gr-qc]} \BibitemShut
  {NoStop}%
\bibitem [{\citenamefont {de~Haro}\ \emph
  {et~al.}(2016{\natexlab{b}})\citenamefont {de~Haro}, \citenamefont
  {Amorós},\ and\ \citenamefont {Pan}}]{deHaro:2016cdm}%
  \BibitemOpen
  \bibfield  {author} {\bibinfo {author} {\bibfnamefont {J.}~\bibnamefont
  {de~Haro}}, \bibinfo {author} {\bibfnamefont {J.}~\bibnamefont {Amorós}}, \
  and\ \bibinfo {author} {\bibfnamefont {S.}~\bibnamefont {Pan}},\ }\href
  {\doibase 10.1103/PhysRevD.94.064060} {\bibfield  {journal} {\bibinfo
  {journal} {Phys. Rev.}\ }\textbf {\bibinfo {volume} {D94}},\ \bibinfo {pages}
  {064060} (\bibinfo {year} {2016}{\natexlab{b}})},\ \Eprint
  {http://arxiv.org/abs/1607.06726} {arXiv:1607.06726 [gr-qc]} \BibitemShut
  {NoStop}%
\bibitem [{\citenamefont {De~Haro}\ and\ \citenamefont
  {Arest\'e~Sal\'o}(2017)}]{deHaro:2017nui}%
  \BibitemOpen
  \bibfield  {author} {\bibinfo {author} {\bibfnamefont {J.}~\bibnamefont
  {De~Haro}}\ and\ \bibinfo {author} {\bibfnamefont {L.}~\bibnamefont
  {Arest\'e~Sal\'o}},\ }\href {\doibase 10.1103/PhysRevD.95.123501} {\bibfield
  {journal} {\bibinfo  {journal} {Phys. Rev. D}\ }\textbf {\bibinfo {volume}
  {95}},\ \bibinfo {pages} {123501} (\bibinfo {year} {2017})},\ \Eprint
  {http://arxiv.org/abs/1702.04212} {arXiv:1702.04212 [gr-qc]} \BibitemShut
  {NoStop}%
\bibitem [{\citenamefont {Arest\'e~Sal\'o}\ and\ \citenamefont
  {de~Haro}(2017)}]{AresteSalo:2017lkv}%
  \BibitemOpen
  \bibfield  {author} {\bibinfo {author} {\bibfnamefont {L.}~\bibnamefont
  {Arest\'e~Sal\'o}}\ and\ \bibinfo {author} {\bibfnamefont {J.}~\bibnamefont
  {de~Haro}},\ }\href {\doibase 10.1140/epjc/s10052-017-5337-0} {\bibfield
  {journal} {\bibinfo  {journal} {Eur. Phys. J. C}\ }\textbf {\bibinfo {volume}
  {77}},\ \bibinfo {pages} {798} (\bibinfo {year} {2017})},\ \Eprint
  {http://arxiv.org/abs/1707.02810} {arXiv:1707.02810 [gr-qc]} \BibitemShut
  {NoStop}%
\bibitem [{\citenamefont {Haro}\ and\ \citenamefont
  {Pan}(2018)}]{Haro:2015ljc}%
  \BibitemOpen
  \bibfield  {author} {\bibinfo {author} {\bibfnamefont {J.}~\bibnamefont
  {Haro}}\ and\ \bibinfo {author} {\bibfnamefont {S.}~\bibnamefont {Pan}},\
  }\href {\doibase 10.1142/S0218271818500529} {\bibfield  {journal} {\bibinfo
  {journal} {Int. J. Mod. Phys.}\ }\textbf {\bibinfo {volume} {D27}},\ \bibinfo
  {pages} {1850052} (\bibinfo {year} {2018})},\ \Eprint
  {http://arxiv.org/abs/1512.03033} {arXiv:1512.03033 [gr-qc]} \BibitemShut
  {NoStop}%
\bibitem [{\citenamefont {Rubio}\ and\ \citenamefont
  {Wetterich}(2017)}]{Rubio:2017gty}%
  \BibitemOpen
  \bibfield  {author} {\bibinfo {author} {\bibfnamefont {J.}~\bibnamefont
  {Rubio}}\ and\ \bibinfo {author} {\bibfnamefont {C.}~\bibnamefont
  {Wetterich}},\ }\href {\doibase 10.1103/PhysRevD.96.063509} {\bibfield
  {journal} {\bibinfo  {journal} {Phys. Rev.}\ }\textbf {\bibinfo {volume}
  {D96}},\ \bibinfo {pages} {063509} (\bibinfo {year} {2017})},\ \Eprint
  {http://arxiv.org/abs/1705.00552} {arXiv:1705.00552 [gr-qc]} \BibitemShut
  {NoStop}%
\bibitem [{\citenamefont {Dimopoulos}\ and\ \citenamefont
  {Markkanen}(2018)}]{Dimopoulos:2018wfg}%
  \BibitemOpen
  \bibfield  {author} {\bibinfo {author} {\bibfnamefont {K.}~\bibnamefont
  {Dimopoulos}}\ and\ \bibinfo {author} {\bibfnamefont {T.}~\bibnamefont
  {Markkanen}},\ }\href {\doibase 10.1088/1475-7516/2018/06/021} {\bibfield
  {journal} {\bibinfo  {journal} {JCAP}\ }\textbf {\bibinfo {volume} {1806}},\
  \bibinfo {pages} {021} (\bibinfo {year} {2018})},\ \Eprint
  {http://arxiv.org/abs/1803.07399} {arXiv:1803.07399 [gr-qc]} \BibitemShut
  {NoStop}%
\bibitem [{\citenamefont {Staicova}\ and\ \citenamefont
  {Stoilov}(2019{\natexlab{a}})}]{Staicova:2018bdy}%
  \BibitemOpen
  \bibfield  {author} {\bibinfo {author} {\bibfnamefont {D.}~\bibnamefont
  {Staicova}}\ and\ \bibinfo {author} {\bibfnamefont {M.}~\bibnamefont
  {Stoilov}},\ }\href {\doibase 10.3390/sym11111387} {\bibfield  {journal}
  {\bibinfo  {journal} {Symmetry}\ }\textbf {\bibinfo {volume} {11}},\ \bibinfo
  {pages} {1387} (\bibinfo {year} {2019}{\natexlab{a}})},\ \Eprint
  {http://arxiv.org/abs/1806.08199} {arXiv:1806.08199 [gr-qc]} \BibitemShut
  {NoStop}%
\bibitem [{\citenamefont {Haro}\ \emph {et~al.}(2020)\citenamefont {Haro},
  \citenamefont {Amorós},\ and\ \citenamefont {Pan}}]{Haro:2019peq}%
  \BibitemOpen
  \bibfield  {author} {\bibinfo {author} {\bibfnamefont {J.}~\bibnamefont
  {Haro}}, \bibinfo {author} {\bibfnamefont {J.}~\bibnamefont {Amorós}}, \
  and\ \bibinfo {author} {\bibfnamefont {S.}~\bibnamefont {Pan}},\ }\href
  {\doibase 10.1140/epjc/s10052-020-7950-6} {\bibfield  {journal} {\bibinfo
  {journal} {Eur. Phys. J.}\ }\textbf {\bibinfo {volume} {C80}},\ \bibinfo
  {pages} {404} (\bibinfo {year} {2020})},\ \Eprint
  {http://arxiv.org/abs/1908.01516} {arXiv:1908.01516 [gr-qc]} \BibitemShut
  {NoStop}%
\bibitem [{\citenamefont {Guendelman}\ \emph {et~al.}(2015)\citenamefont
  {Guendelman}, \citenamefont {Herrera}, \citenamefont {Labrana}, \citenamefont
  {Nissimov},\ and\ \citenamefont {Pacheva}}]{Guendelman:2014bva}%
  \BibitemOpen
  \bibfield  {author} {\bibinfo {author} {\bibfnamefont {E.}~\bibnamefont
  {Guendelman}}, \bibinfo {author} {\bibfnamefont {R.}~\bibnamefont {Herrera}},
  \bibinfo {author} {\bibfnamefont {P.}~\bibnamefont {Labrana}}, \bibinfo
  {author} {\bibfnamefont {E.}~\bibnamefont {Nissimov}}, \ and\ \bibinfo
  {author} {\bibfnamefont {S.}~\bibnamefont {Pacheva}},\ }\href {\doibase
  10.1007/s10714-015-1852-1} {\bibfield  {journal} {\bibinfo  {journal} {Gen.
  Rel. Grav.}\ }\textbf {\bibinfo {volume} {47}},\ \bibinfo {pages} {10}
  (\bibinfo {year} {2015})},\ \Eprint {http://arxiv.org/abs/1408.5344}
  {arXiv:1408.5344 [gr-qc]} \BibitemShut {NoStop}%
\bibitem [{\citenamefont {Guendelman}\ and\ \citenamefont
  {Herrera}(2016)}]{Guendelman:2015liz}%
  \BibitemOpen
  \bibfield  {author} {\bibinfo {author} {\bibfnamefont {E.~I.}\ \bibnamefont
  {Guendelman}}\ and\ \bibinfo {author} {\bibfnamefont {R.}~\bibnamefont
  {Herrera}},\ }\href {\doibase 10.1007/s10714-015-1999-9} {\bibfield
  {journal} {\bibinfo  {journal} {Gen. Rel. Grav.}\ }\textbf {\bibinfo {volume}
  {48}},\ \bibinfo {pages} {3} (\bibinfo {year} {2016})},\ \Eprint
  {http://arxiv.org/abs/1511.08645} {arXiv:1511.08645 [gr-qc]} \BibitemShut
  {NoStop}%
\bibitem [{\citenamefont {van~de Bruck}\ \emph {et~al.}(2017)\citenamefont
  {van~de Bruck}, \citenamefont {Dimopoulos}, \citenamefont {Longden},\ and\
  \citenamefont {Owen}}]{vandeBruck:2017voa}%
  \BibitemOpen
  \bibfield  {author} {\bibinfo {author} {\bibfnamefont {C.}~\bibnamefont
  {van~de Bruck}}, \bibinfo {author} {\bibfnamefont {K.}~\bibnamefont
  {Dimopoulos}}, \bibinfo {author} {\bibfnamefont {C.}~\bibnamefont {Longden}},
  \ and\ \bibinfo {author} {\bibfnamefont {C.}~\bibnamefont {Owen}},\
  }\href@noop {} {\  (\bibinfo {year} {2017})},\ \Eprint
  {http://arxiv.org/abs/1707.06839} {arXiv:1707.06839 [astro-ph.CO]}
  \BibitemShut {NoStop}%
\bibitem [{\citenamefont {Dimopoulos}\ \emph {et~al.}(2019)\citenamefont
  {Dimopoulos}, \citenamefont {Kar\v{c}iauskas},\ and\ \citenamefont
  {Owen}}]{Dimopoulos:2019ogl}%
  \BibitemOpen
  \bibfield  {author} {\bibinfo {author} {\bibfnamefont {K.}~\bibnamefont
  {Dimopoulos}}, \bibinfo {author} {\bibfnamefont {M.}~\bibnamefont
  {Kar\v{c}iauskas}}, \ and\ \bibinfo {author} {\bibfnamefont {C.}~\bibnamefont
  {Owen}},\ }\href {\doibase 10.1103/PhysRevD.100.083530} {\bibfield  {journal}
  {\bibinfo  {journal} {Phys. Rev. D}\ }\textbf {\bibinfo {volume} {100}},\
  \bibinfo {pages} {083530} (\bibinfo {year} {2019})},\ \Eprint
  {http://arxiv.org/abs/1907.04676} {arXiv:1907.04676 [hep-ph]} \BibitemShut
  {NoStop}%
\bibitem [{\citenamefont {Kleidis}\ and\ \citenamefont
  {Oikonomou}(2019)}]{Kleidis:2019ywv}%
  \BibitemOpen
  \bibfield  {author} {\bibinfo {author} {\bibfnamefont {K.}~\bibnamefont
  {Kleidis}}\ and\ \bibinfo {author} {\bibfnamefont {V.~K.}\ \bibnamefont
  {Oikonomou}},\ }\href {\doibase 10.1016/j.nuclphysb.2019.114765} {\bibfield
  {journal} {\bibinfo  {journal} {Nucl. Phys. B}\ }\textbf {\bibinfo {volume}
  {948}},\ \bibinfo {pages} {114765} (\bibinfo {year} {2019})},\ \Eprint
  {http://arxiv.org/abs/1909.05318} {arXiv:1909.05318 [gr-qc]} \BibitemShut
  {NoStop}%
\bibitem [{\citenamefont {Lima}\ and\ \citenamefont
  {Ramos}(2019)}]{Lima:2019yyv}%
  \BibitemOpen
  \bibfield  {author} {\bibinfo {author} {\bibfnamefont {G.~B.~F.}\
  \bibnamefont {Lima}}\ and\ \bibinfo {author} {\bibfnamefont {R.~O.}\
  \bibnamefont {Ramos}},\ }\href {\doibase 10.1103/PhysRevD.100.123529}
  {\bibfield  {journal} {\bibinfo  {journal} {Phys. Rev. D}\ }\textbf {\bibinfo
  {volume} {100}},\ \bibinfo {pages} {123529} (\bibinfo {year} {2019})},\
  \Eprint {http://arxiv.org/abs/1910.05185} {arXiv:1910.05185 [astro-ph.CO]}
  \BibitemShut {NoStop}%
\bibitem [{\citenamefont {Staicova}(2020)}]{Staicova:2020wph}%
  \BibitemOpen
  \bibfield  {author} {\bibinfo {author} {\bibfnamefont {D.}~\bibnamefont
  {Staicova}},\ }\href@noop {} {\  (\bibinfo {year} {2020})},\ \Eprint
  {http://arxiv.org/abs/2011.02967} {arXiv:2011.02967 [gr-qc]} \BibitemShut
  {NoStop}%
\bibitem [{\citenamefont {Benisty}\ \emph {et~al.}(2020)\citenamefont
  {Benisty}, \citenamefont {Guendelman},\ and\ \citenamefont
  {Saridakis}}]{Benisty:2019vej}%
  \BibitemOpen
  \bibfield  {author} {\bibinfo {author} {\bibfnamefont {D.}~\bibnamefont
  {Benisty}}, \bibinfo {author} {\bibfnamefont {E.~I.}\ \bibnamefont
  {Guendelman}}, \ and\ \bibinfo {author} {\bibfnamefont {E.~N.}\ \bibnamefont
  {Saridakis}},\ }\href {\doibase 10.1140/epjc/s10052-020-8054-z} {\bibfield
  {journal} {\bibinfo  {journal} {Eur. Phys. J.}\ }\textbf {\bibinfo {volume}
  {C80}},\ \bibinfo {pages} {480} (\bibinfo {year} {2020})},\ \Eprint
  {http://arxiv.org/abs/1909.01982} {arXiv:1909.01982 [gr-qc]} \BibitemShut
  {NoStop}%
\bibitem [{\citenamefont {Rosa}\ and\ \citenamefont
  {Ventura}(2019)}]{Rosa:2019jci}%
  \BibitemOpen
  \bibfield  {author} {\bibinfo {author} {\bibfnamefont {J.~a.~G.}\
  \bibnamefont {Rosa}}\ and\ \bibinfo {author} {\bibfnamefont {L.~B.}\
  \bibnamefont {Ventura}},\ }\href {\doibase 10.1016/j.physletb.2019.134984}
  {\bibfield  {journal} {\bibinfo  {journal} {Phys. Lett. B}\ }\textbf
  {\bibinfo {volume} {798}},\ \bibinfo {pages} {134984} (\bibinfo {year}
  {2019})},\ \Eprint {http://arxiv.org/abs/1906.11835} {arXiv:1906.11835
  [hep-ph]} \BibitemShut {NoStop}%
\bibitem [{\citenamefont {Staicova}\ and\ \citenamefont
  {Stoilov}(2019{\natexlab{b}})}]{Staicova:2019ksr}%
  \BibitemOpen
  \bibfield  {author} {\bibinfo {author} {\bibfnamefont {D.}~\bibnamefont
  {Staicova}}\ and\ \bibinfo {author} {\bibfnamefont {M.}~\bibnamefont
  {Stoilov}},\ }\href {\doibase 10.1142/S0217751X19500994} {\bibfield
  {journal} {\bibinfo  {journal} {Int. J. Mod. Phys. A}\ }\textbf {\bibinfo
  {volume} {34}},\ \bibinfo {pages} {1950099} (\bibinfo {year}
  {2019}{\natexlab{b}})},\ \Eprint {http://arxiv.org/abs/1906.08516}
  {arXiv:1906.08516 [gr-qc]} \BibitemShut {NoStop}%
\bibitem [{\citenamefont {Dimopoulos}\ and\ \citenamefont
  {S\'anchez~L\'opez}(2020)}]{Dimopoulos:2020pas}%
  \BibitemOpen
  \bibfield  {author} {\bibinfo {author} {\bibfnamefont {K.}~\bibnamefont
  {Dimopoulos}}\ and\ \bibinfo {author} {\bibfnamefont {S.}~\bibnamefont
  {S\'anchez~L\'opez}},\ }\href@noop {} {\  (\bibinfo {year} {2020})},\ \Eprint
  {http://arxiv.org/abs/2012.06831} {arXiv:2012.06831 [gr-qc]} \BibitemShut
  {NoStop}%
\bibitem [{\citenamefont {Es-haghi}\ and\ \citenamefont
  {Sheykhi}(2020)}]{Es-haghi:2020oab}%
  \BibitemOpen
  \bibfield  {author} {\bibinfo {author} {\bibfnamefont {M.}~\bibnamefont
  {Es-haghi}}\ and\ \bibinfo {author} {\bibfnamefont {A.}~\bibnamefont
  {Sheykhi}},\ }\href@noop {} {\  (\bibinfo {year} {2020})},\ \Eprint
  {http://arxiv.org/abs/2012.08035} {arXiv:2012.08035 [astro-ph.CO]}
  \BibitemShut {NoStop}%
\bibitem [{\citenamefont {Banerjee}\ \emph {et~al.}(2020)\citenamefont
  {Banerjee}, \citenamefont {Cai}, \citenamefont {Heisenberg}, \citenamefont
  {Colg\'ain}, \citenamefont {Sheikh-Jabbari},\ and\ \citenamefont
  {Yang}}]{Banerjee:2020xcn}%
  \BibitemOpen
  \bibfield  {author} {\bibinfo {author} {\bibfnamefont {A.}~\bibnamefont
  {Banerjee}}, \bibinfo {author} {\bibfnamefont {H.}~\bibnamefont {Cai}},
  \bibinfo {author} {\bibfnamefont {L.}~\bibnamefont {Heisenberg}}, \bibinfo
  {author} {\bibfnamefont {E.~O.}\ \bibnamefont {Colg\'ain}}, \bibinfo {author}
  {\bibfnamefont {M.~M.}\ \bibnamefont {Sheikh-Jabbari}}, \ and\ \bibinfo
  {author} {\bibfnamefont {T.}~\bibnamefont {Yang}},\ }\href@noop {} {\
  (\bibinfo {year} {2020})},\ \Eprint {http://arxiv.org/abs/2006.00244}
  {arXiv:2006.00244 [astro-ph.CO]} \BibitemShut {NoStop}%
\bibitem [{\citenamefont {Rodrigues}\ \emph
  {et~al.}(2020{\natexlab{a}})\citenamefont {Rodrigues}, \citenamefont
  {Benetti}, \citenamefont {Campista},\ and\ \citenamefont
  {Alcaniz}}]{Rodrigues:2020dod}%
  \BibitemOpen
  \bibfield  {author} {\bibinfo {author} {\bibfnamefont {J.~G.}\ \bibnamefont
  {Rodrigues}}, \bibinfo {author} {\bibfnamefont {M.}~\bibnamefont {Benetti}},
  \bibinfo {author} {\bibfnamefont {M.}~\bibnamefont {Campista}}, \ and\
  \bibinfo {author} {\bibfnamefont {J.}~\bibnamefont {Alcaniz}},\ }\href
  {\doibase 10.1088/1475-7516/2020/07/007} {\bibfield  {journal} {\bibinfo
  {journal} {JCAP}\ }\textbf {\bibinfo {volume} {07}},\ \bibinfo {pages} {007}
  (\bibinfo {year} {2020}{\natexlab{a}})},\ \Eprint
  {http://arxiv.org/abs/2002.05154} {arXiv:2002.05154 [astro-ph.CO]}
  \BibitemShut {NoStop}%
\bibitem [{\citenamefont {Ade}\ \emph {et~al.}(2016)\citenamefont {Ade} \emph
  {et~al.}}]{Ade:2015xua}%
  \BibitemOpen
  \bibfield  {author} {\bibinfo {author} {\bibfnamefont {P.~A.~R.}\
  \bibnamefont {Ade}} \emph {et~al.} (\bibinfo {collaboration} {Planck}),\
  }\href {\doibase 10.1051/0004-6361/201525830} {\bibfield  {journal} {\bibinfo
   {journal} {Astron. Astrophys.}\ }\textbf {\bibinfo {volume} {594}},\
  \bibinfo {pages} {A13} (\bibinfo {year} {2016})},\ \Eprint
  {http://arxiv.org/abs/1502.01589} {arXiv:1502.01589 [astro-ph.CO]}
  \BibitemShut {NoStop}%
\bibitem [{\citenamefont {Akrami}\ \emph
  {et~al.}(2018{\natexlab{a}})\citenamefont {Akrami} \emph
  {et~al.}}]{Akrami:2018odb}%
  \BibitemOpen
  \bibfield  {author} {\bibinfo {author} {\bibfnamefont {Y.}~\bibnamefont
  {Akrami}} \emph {et~al.} (\bibinfo {collaboration} {Planck}),\ }\href@noop {}
  {\  (\bibinfo {year} {2018}{\natexlab{a}})},\ \Eprint
  {http://arxiv.org/abs/1807.06211} {arXiv:1807.06211 [astro-ph.CO]}
  \BibitemShut {NoStop}%
\bibitem [{\citenamefont {Aghanim}\ \emph {et~al.}(2018)\citenamefont {Aghanim}
  \emph {et~al.}}]{Aghanim:2018eyx}%
  \BibitemOpen
  \bibfield  {author} {\bibinfo {author} {\bibfnamefont {N.}~\bibnamefont
  {Aghanim}} \emph {et~al.} (\bibinfo {collaboration} {Planck}),\ }\href@noop
  {} {\  (\bibinfo {year} {2018})},\ \Eprint {http://arxiv.org/abs/1807.06209}
  {arXiv:1807.06209 [astro-ph.CO]} \BibitemShut {NoStop}%
\bibitem [{\citenamefont {Akrami}\ \emph
  {et~al.}(2018{\natexlab{b}})\citenamefont {Akrami}, \citenamefont {Kallosh},
  \citenamefont {Linde},\ and\ \citenamefont {Vardanyan}}]{Akrami:2017cir}%
  \BibitemOpen
  \bibfield  {author} {\bibinfo {author} {\bibfnamefont {Y.}~\bibnamefont
  {Akrami}}, \bibinfo {author} {\bibfnamefont {R.}~\bibnamefont {Kallosh}},
  \bibinfo {author} {\bibfnamefont {A.}~\bibnamefont {Linde}}, \ and\ \bibinfo
  {author} {\bibfnamefont {V.}~\bibnamefont {Vardanyan}},\ }\href {\doibase
  10.1088/1475-7516/2018/06/041} {\bibfield  {journal} {\bibinfo  {journal}
  {JCAP}\ }\textbf {\bibinfo {volume} {06}},\ \bibinfo {pages} {041} (\bibinfo
  {year} {2018}{\natexlab{b}})},\ \Eprint {http://arxiv.org/abs/1712.09693}
  {arXiv:1712.09693 [hep-th]} \BibitemShut {NoStop}%
\bibitem [{\citenamefont {Akrami}\ \emph {et~al.}(2020)\citenamefont {Akrami},
  \citenamefont {Casas}, \citenamefont {Deng},\ and\ \citenamefont
  {Vardanyan}}]{Akrami:2020zxw}%
  \BibitemOpen
  \bibfield  {author} {\bibinfo {author} {\bibfnamefont {Y.}~\bibnamefont
  {Akrami}}, \bibinfo {author} {\bibfnamefont {S.}~\bibnamefont {Casas}},
  \bibinfo {author} {\bibfnamefont {S.}~\bibnamefont {Deng}}, \ and\ \bibinfo
  {author} {\bibfnamefont {V.}~\bibnamefont {Vardanyan}},\ }\href@noop {} {\
  (\bibinfo {year} {2020})},\ \Eprint {http://arxiv.org/abs/2010.15822}
  {arXiv:2010.15822 [astro-ph.CO]} \BibitemShut {NoStop}%
\bibitem [{\citenamefont {Rodrigues}\ \emph
  {et~al.}(2020{\natexlab{b}})\citenamefont {Rodrigues}, \citenamefont
  {Santos~da Costa},\ and\ \citenamefont {Alcaniz}}]{Rodrigues:2020jsv}%
  \BibitemOpen
  \bibfield  {author} {\bibinfo {author} {\bibfnamefont {J.~G.}\ \bibnamefont
  {Rodrigues}}, \bibinfo {author} {\bibfnamefont {S.}~\bibnamefont {Santos~da
  Costa}}, \ and\ \bibinfo {author} {\bibfnamefont {J.~S.}\ \bibnamefont
  {Alcaniz}},\ }\href@noop {} {\  (\bibinfo {year} {2020}{\natexlab{b}})},\
  \Eprint {http://arxiv.org/abs/2007.10763} {arXiv:2007.10763 [astro-ph.CO]}
  \BibitemShut {NoStop}%
\bibitem [{\citenamefont {Elizalde}\ \emph {et~al.}(2016)\citenamefont
  {Elizalde}, \citenamefont {Odintsov}, \citenamefont {Pozdeeva},\ and\
  \citenamefont {Vernov}}]{Elizalde:2015nya}%
  \BibitemOpen
  \bibfield  {author} {\bibinfo {author} {\bibfnamefont {E.}~\bibnamefont
  {Elizalde}}, \bibinfo {author} {\bibfnamefont {S.~D.}\ \bibnamefont
  {Odintsov}}, \bibinfo {author} {\bibfnamefont {E.~O.}\ \bibnamefont
  {Pozdeeva}}, \ and\ \bibinfo {author} {\bibfnamefont {S.~{\relax Yu}.}\
  \bibnamefont {Vernov}},\ }\href {\doibase 10.1088/1475-7516/2016/02/025}
  {\bibfield  {journal} {\bibinfo  {journal} {JCAP}\ }\textbf {\bibinfo
  {volume} {1602}},\ \bibinfo {pages} {025} (\bibinfo {year} {2016})},\ \Eprint
  {http://arxiv.org/abs/1509.08817} {arXiv:1509.08817 [gr-qc]} \BibitemShut
  {NoStop}%
\bibitem [{\citenamefont {Dubinin}\ \emph {et~al.}(2018)\citenamefont
  {Dubinin}, \citenamefont {Petrova}, \citenamefont {Pozdeeva},\ and\
  \citenamefont {Vernov}}]{Dubinin:2017irq}%
  \BibitemOpen
  \bibfield  {author} {\bibinfo {author} {\bibfnamefont {M.~N.}\ \bibnamefont
  {Dubinin}}, \bibinfo {author} {\bibfnamefont {E.~{\relax Yu}.}\ \bibnamefont
  {Petrova}}, \bibinfo {author} {\bibfnamefont {E.~O.}\ \bibnamefont
  {Pozdeeva}}, \ and\ \bibinfo {author} {\bibfnamefont {S.~{\relax Yu}.}\
  \bibnamefont {Vernov}},\ }\href {\doibase 10.1142/S0219887818400017}
  {\bibfield  {journal} {\bibinfo  {journal} {Int. J. Geom. Meth. Mod. Phys.}\
  }\textbf {\bibinfo {volume} {15}},\ \bibinfo {pages} {1840001} (\bibinfo
  {year} {2018})},\ \Eprint {http://arxiv.org/abs/1712.03072} {arXiv:1712.03072
  [hep-ph]} \BibitemShut {NoStop}%
\bibitem [{\citenamefont {Pozdeeva}(2020)}]{Pozdeeva:2020shl}%
  \BibitemOpen
  \bibfield  {author} {\bibinfo {author} {\bibfnamefont {E.~O.}\ \bibnamefont
  {Pozdeeva}},\ }\href {\doibase 10.1140/epjc/s10052-020-8176-3} {\bibfield
  {journal} {\bibinfo  {journal} {Eur. Phys. J.}\ }\textbf {\bibinfo {volume}
  {C80}},\ \bibinfo {pages} {612} (\bibinfo {year} {2020})},\ \Eprint
  {http://arxiv.org/abs/2005.10133} {arXiv:2005.10133 [gr-qc]} \BibitemShut
  {NoStop}%
\bibitem [{\citenamefont {Minkowski}(1977)}]{Minkowski:1977sc}%
  \BibitemOpen
  \bibfield  {author} {\bibinfo {author} {\bibfnamefont {P.}~\bibnamefont
  {Minkowski}},\ }\href {\doibase 10.1016/0370-2693(77)90435-X} {\bibfield
  {journal} {\bibinfo  {journal} {Phys. Lett.}\ }\textbf {\bibinfo {volume}
  {67B}},\ \bibinfo {pages} {421} (\bibinfo {year} {1977})}\BibitemShut
  {NoStop}%
\bibitem [{\citenamefont {Yanagida}(1980)}]{Yanagida:1980xy}%
  \BibitemOpen
  \bibfield  {author} {\bibinfo {author} {\bibfnamefont {T.}~\bibnamefont
  {Yanagida}},\ }\href {\doibase 10.1143/PTP.64.1103} {\bibfield  {journal}
  {\bibinfo  {journal} {Prog. Theor. Phys.}\ }\textbf {\bibinfo {volume}
  {64}},\ \bibinfo {pages} {1103} (\bibinfo {year} {1980})}\BibitemShut
  {NoStop}%
\bibitem [{\citenamefont {Schechter}\ and\ \citenamefont
  {Valle}(1980)}]{Schechter:1980gr}%
  \BibitemOpen
  \bibfield  {author} {\bibinfo {author} {\bibfnamefont {J.}~\bibnamefont
  {Schechter}}\ and\ \bibinfo {author} {\bibfnamefont {J.~W.~F.}\ \bibnamefont
  {Valle}},\ }\href {\doibase 10.1103/PhysRevD.22.2227} {\bibfield  {journal}
  {\bibinfo  {journal} {Phys. Rev.}\ }\textbf {\bibinfo {volume} {D22}},\
  \bibinfo {pages} {2227} (\bibinfo {year} {1980})}\BibitemShut {NoStop}%
\bibitem [{\citenamefont {Davidson}\ and\ \citenamefont
  {Wali}(1988)}]{Davidson:1987tr}%
  \BibitemOpen
  \bibfield  {author} {\bibinfo {author} {\bibfnamefont {A.}~\bibnamefont
  {Davidson}}\ and\ \bibinfo {author} {\bibfnamefont {K.~C.}\ \bibnamefont
  {Wali}},\ }\href {\doibase 10.1103/PhysRevLett.60.1813} {\bibfield  {journal}
  {\bibinfo  {journal} {Phys. Rev. Lett.}\ }\textbf {\bibinfo {volume} {60}},\
  \bibinfo {pages} {1813} (\bibinfo {year} {1988})}\BibitemShut {NoStop}%
\bibitem [{\citenamefont {Davidson}\ and\ \citenamefont
  {Wali}(1987)}]{Davidson:1987mh}%
  \BibitemOpen
  \bibfield  {author} {\bibinfo {author} {\bibfnamefont {A.}~\bibnamefont
  {Davidson}}\ and\ \bibinfo {author} {\bibfnamefont {K.~C.}\ \bibnamefont
  {Wali}},\ }\href {\doibase 10.1103/PhysRevLett.59.393} {\bibfield  {journal}
  {\bibinfo  {journal} {Phys. Rev. Lett.}\ }\textbf {\bibinfo {volume} {59}},\
  \bibinfo {pages} {393} (\bibinfo {year} {1987})}\BibitemShut {NoStop}%
\bibitem [{\citenamefont {Rajpoot}(1987)}]{Rajpoot:1987ji}%
  \BibitemOpen
  \bibfield  {author} {\bibinfo {author} {\bibfnamefont {S.}~\bibnamefont
  {Rajpoot}},\ }\href {\doibase 10.1103/PhysRevD.36.1479} {\bibfield  {journal}
  {\bibinfo  {journal} {Phys. Rev.}\ }\textbf {\bibinfo {volume} {D36}},\
  \bibinfo {pages} {1479} (\bibinfo {year} {1987})}\BibitemShut {NoStop}%
\bibitem [{\citenamefont {Guendelman}(1999)}]{Guendelman:1999qt}%
  \BibitemOpen
  \bibfield  {author} {\bibinfo {author} {\bibfnamefont {E.~I.}\ \bibnamefont
  {Guendelman}},\ }\href {\doibase 10.1142/S0217732399001103} {\bibfield
  {journal} {\bibinfo  {journal} {Mod. Phys. Lett.}\ }\textbf {\bibinfo
  {volume} {A14}},\ \bibinfo {pages} {1043} (\bibinfo {year} {1999})},\ \Eprint
  {http://arxiv.org/abs/gr-qc/9901017} {arXiv:gr-qc/9901017 [gr-qc]}
  \BibitemShut {NoStop}%
\bibitem [{\citenamefont {Guendelman}(2001)}]{Guendelman:2001bu}%
  \BibitemOpen
  \bibfield  {author} {\bibinfo {author} {\bibfnamefont {E.~I.}\ \bibnamefont
  {Guendelman}},\ }in\ \href@noop {} {\emph {\bibinfo {booktitle} {{8th
  International Symposium on Particles Strings and Cosmology}}}}\ (\bibinfo
  {year} {2001})\ \Eprint {http://arxiv.org/abs/hep-th/0106085}
  {arXiv:hep-th/0106085} \BibitemShut {NoStop}%
\bibitem [{\citenamefont {Guendelman}\ and\ \citenamefont
  {Katz}(2003)}]{Guendelman:2002js}%
  \BibitemOpen
  \bibfield  {author} {\bibinfo {author} {\bibfnamefont {E.~I.}\ \bibnamefont
  {Guendelman}}\ and\ \bibinfo {author} {\bibfnamefont {O.}~\bibnamefont
  {Katz}},\ }\href {\doibase 10.1088/0264-9381/20/9/309} {\bibfield  {journal}
  {\bibinfo  {journal} {Class. Quant. Grav.}\ }\textbf {\bibinfo {volume}
  {20}},\ \bibinfo {pages} {1715} (\bibinfo {year} {2003})},\ \Eprint
  {http://arxiv.org/abs/gr-qc/0211095} {arXiv:gr-qc/0211095} \BibitemShut
  {NoStop}%
\bibitem [{\citenamefont {Martin}\ \emph {et~al.}(2016)\citenamefont {Martin},
  \citenamefont {Ringeval},\ and\ \citenamefont {Vennin}}]{Martin:2016iqo}%
  \BibitemOpen
  \bibfield  {author} {\bibinfo {author} {\bibfnamefont {J.}~\bibnamefont
  {Martin}}, \bibinfo {author} {\bibfnamefont {C.}~\bibnamefont {Ringeval}}, \
  and\ \bibinfo {author} {\bibfnamefont {V.}~\bibnamefont {Vennin}},\ }\href
  {\doibase 10.1103/PhysRevD.94.123521} {\bibfield  {journal} {\bibinfo
  {journal} {Phys. Rev. D}\ }\textbf {\bibinfo {volume} {94}},\ \bibinfo
  {pages} {123521} (\bibinfo {year} {2016})},\ \Eprint
  {http://arxiv.org/abs/1609.04739} {arXiv:1609.04739 [astro-ph.CO]}
  \BibitemShut {NoStop}%
\bibitem [{\citenamefont {Martin}\ and\ \citenamefont
  {Ringeval}(2010)}]{Martin:2010kz}%
  \BibitemOpen
  \bibfield  {author} {\bibinfo {author} {\bibfnamefont {J.}~\bibnamefont
  {Martin}}\ and\ \bibinfo {author} {\bibfnamefont {C.}~\bibnamefont
  {Ringeval}},\ }\href {\doibase 10.1103/PhysRevD.82.023511} {\bibfield
  {journal} {\bibinfo  {journal} {Phys. Rev.}\ }\textbf {\bibinfo {volume}
  {D82}},\ \bibinfo {pages} {023511} (\bibinfo {year} {2010})},\ \Eprint
  {http://arxiv.org/abs/1004.5525} {arXiv:1004.5525 [astro-ph.CO]} \BibitemShut
  {NoStop}%
\bibitem [{\citenamefont {Martin}\ \emph {et~al.}(2015)\citenamefont {Martin},
  \citenamefont {Ringeval},\ and\ \citenamefont {Vennin}}]{Martin:2014nya}%
  \BibitemOpen
  \bibfield  {author} {\bibinfo {author} {\bibfnamefont {J.}~\bibnamefont
  {Martin}}, \bibinfo {author} {\bibfnamefont {C.}~\bibnamefont {Ringeval}}, \
  and\ \bibinfo {author} {\bibfnamefont {V.}~\bibnamefont {Vennin}},\ }\href
  {\doibase 10.1103/PhysRevLett.114.081303} {\bibfield  {journal} {\bibinfo
  {journal} {Phys. Rev. Lett.}\ }\textbf {\bibinfo {volume} {114}},\ \bibinfo
  {pages} {081303} (\bibinfo {year} {2015})},\ \Eprint
  {http://arxiv.org/abs/1410.7958} {arXiv:1410.7958 [astro-ph.CO]} \BibitemShut
  {NoStop}%
\bibitem [{\citenamefont {Joyce}(1997)}]{Joyce:1996cp}%
  \BibitemOpen
  \bibfield  {author} {\bibinfo {author} {\bibfnamefont {M.}~\bibnamefont
  {Joyce}},\ }\href {\doibase 10.1103/PhysRevD.55.1875} {\bibfield  {journal}
  {\bibinfo  {journal} {Phys. Rev. D}\ }\textbf {\bibinfo {volume} {55}},\
  \bibinfo {pages} {1875} (\bibinfo {year} {1997})},\ \Eprint
  {http://arxiv.org/abs/hep-ph/9606223} {arXiv:hep-ph/9606223} \BibitemShut
  {NoStop}%
\bibitem [{\citenamefont {Spokoiny}(1993)}]{Spokoiny:1993kt}%
  \BibitemOpen
  \bibfield  {author} {\bibinfo {author} {\bibfnamefont {B.}~\bibnamefont
  {Spokoiny}},\ }\href {\doibase 10.1016/0370-2693(93)90155-B} {\bibfield
  {journal} {\bibinfo  {journal} {Phys. Lett. B}\ }\textbf {\bibinfo {volume}
  {315}},\ \bibinfo {pages} {40} (\bibinfo {year} {1993})},\ \Eprint
  {http://arxiv.org/abs/gr-qc/9306008} {arXiv:gr-qc/9306008} \BibitemShut
  {NoStop}%
\bibitem [{\citenamefont {Rehagen}\ and\ \citenamefont
  {Gelmini}(2015)}]{Rehagen:2015zma}%
  \BibitemOpen
  \bibfield  {author} {\bibinfo {author} {\bibfnamefont {T.}~\bibnamefont
  {Rehagen}}\ and\ \bibinfo {author} {\bibfnamefont {G.~B.}\ \bibnamefont
  {Gelmini}},\ }\href {\doibase 10.1088/1475-7516/2015/06/039} {\bibfield
  {journal} {\bibinfo  {journal} {JCAP}\ }\textbf {\bibinfo {volume} {06}},\
  \bibinfo {pages} {039} (\bibinfo {year} {2015})},\ \Eprint
  {http://arxiv.org/abs/1504.03768} {arXiv:1504.03768 [hep-ph]} \BibitemShut
  {NoStop}%
\bibitem [{\citenamefont {Ford}(1987)}]{Ford:1986sy}%
  \BibitemOpen
  \bibfield  {author} {\bibinfo {author} {\bibfnamefont {L.~H.}\ \bibnamefont
  {Ford}},\ }\href {\doibase 10.1103/PhysRevD.35.2955} {\bibfield  {journal}
  {\bibinfo  {journal} {Phys. Rev.}\ }\textbf {\bibinfo {volume} {D35}},\
  \bibinfo {pages} {2955} (\bibinfo {year} {1987})}\BibitemShut {NoStop}%
\bibitem [{\citenamefont {Haro}\ \emph
  {et~al.}(2019{\natexlab{b}})\citenamefont {Haro}, \citenamefont {Yang},\ and\
  \citenamefont {Pan}}]{Haro:2018zdb}%
  \BibitemOpen
  \bibfield  {author} {\bibinfo {author} {\bibfnamefont {J.}~\bibnamefont
  {Haro}}, \bibinfo {author} {\bibfnamefont {W.}~\bibnamefont {Yang}}, \ and\
  \bibinfo {author} {\bibfnamefont {S.}~\bibnamefont {Pan}},\ }\href {\doibase
  10.1088/1475-7516/2019/01/023} {\bibfield  {journal} {\bibinfo  {journal}
  {JCAP}\ }\textbf {\bibinfo {volume} {01}},\ \bibinfo {pages} {023} (\bibinfo
  {year} {2019}{\natexlab{b}})},\ \Eprint {http://arxiv.org/abs/1811.07371}
  {arXiv:1811.07371 [gr-qc]} \BibitemShut {NoStop}%
\bibitem [{\citenamefont {Hashiba}\ and\ \citenamefont
  {Yokoyama}(2019)}]{Hashiba:2018iff}%
  \BibitemOpen
  \bibfield  {author} {\bibinfo {author} {\bibfnamefont {S.}~\bibnamefont
  {Hashiba}}\ and\ \bibinfo {author} {\bibfnamefont {J.}~\bibnamefont
  {Yokoyama}},\ }\href {\doibase 10.1088/1475-7516/2019/01/028} {\bibfield
  {journal} {\bibinfo  {journal} {JCAP}\ }\textbf {\bibinfo {volume} {01}},\
  \bibinfo {pages} {028} (\bibinfo {year} {2019})},\ \Eprint
  {http://arxiv.org/abs/1809.05410} {arXiv:1809.05410 [gr-qc]} \BibitemShut
  {NoStop}%
\bibitem [{\citenamefont {Chung}\ \emph {et~al.}(1998)\citenamefont {Chung},
  \citenamefont {Kolb},\ and\ \citenamefont {Riotto}}]{Chung:1998zb}%
  \BibitemOpen
  \bibfield  {author} {\bibinfo {author} {\bibfnamefont {D.~J.~H.}\
  \bibnamefont {Chung}}, \bibinfo {author} {\bibfnamefont {E.~W.}\ \bibnamefont
  {Kolb}}, \ and\ \bibinfo {author} {\bibfnamefont {A.}~\bibnamefont
  {Riotto}},\ }\href {\doibase 10.1103/PhysRevD.59.023501} {\bibfield
  {journal} {\bibinfo  {journal} {Phys. Rev. D}\ }\textbf {\bibinfo {volume}
  {59}},\ \bibinfo {pages} {023501} (\bibinfo {year} {1998})},\ \Eprint
  {http://arxiv.org/abs/hep-ph/9802238} {arXiv:hep-ph/9802238} \BibitemShut
  {NoStop}%
\bibitem [{\citenamefont {Chung}\ \emph {et~al.}(2001)\citenamefont {Chung},
  \citenamefont {Crotty}, \citenamefont {Kolb},\ and\ \citenamefont
  {Riotto}}]{Chung:2001cb}%
  \BibitemOpen
  \bibfield  {author} {\bibinfo {author} {\bibfnamefont {D.~J.~H.}\
  \bibnamefont {Chung}}, \bibinfo {author} {\bibfnamefont {P.}~\bibnamefont
  {Crotty}}, \bibinfo {author} {\bibfnamefont {E.~W.}\ \bibnamefont {Kolb}}, \
  and\ \bibinfo {author} {\bibfnamefont {A.}~\bibnamefont {Riotto}},\ }\href
  {\doibase 10.1103/PhysRevD.64.043503} {\bibfield  {journal} {\bibinfo
  {journal} {Phys. Rev. D}\ }\textbf {\bibinfo {volume} {64}},\ \bibinfo
  {pages} {043503} (\bibinfo {year} {2001})},\ \Eprint
  {http://arxiv.org/abs/hep-ph/0104100} {arXiv:hep-ph/0104100} \BibitemShut
  {NoStop}%
\bibitem [{\citenamefont {Felder}\ \emph
  {et~al.}(1999{\natexlab{a}})\citenamefont {Felder}, \citenamefont {Kofman},\
  and\ \citenamefont {Linde}}]{Felder:1998vq}%
  \BibitemOpen
  \bibfield  {author} {\bibinfo {author} {\bibfnamefont {G.~N.}\ \bibnamefont
  {Felder}}, \bibinfo {author} {\bibfnamefont {L.}~\bibnamefont {Kofman}}, \
  and\ \bibinfo {author} {\bibfnamefont {A.~D.}\ \bibnamefont {Linde}},\ }\href
  {\doibase 10.1103/PhysRevD.59.123523} {\bibfield  {journal} {\bibinfo
  {journal} {Phys. Rev. D}\ }\textbf {\bibinfo {volume} {59}},\ \bibinfo
  {pages} {123523} (\bibinfo {year} {1999}{\natexlab{a}})},\ \Eprint
  {http://arxiv.org/abs/hep-ph/9812289} {arXiv:hep-ph/9812289} \BibitemShut
  {NoStop}%
\bibitem [{\citenamefont {Felder}\ \emph
  {et~al.}(1999{\natexlab{b}})\citenamefont {Felder}, \citenamefont {Kofman},\
  and\ \citenamefont {Linde}}]{Felder:1999pv}%
  \BibitemOpen
  \bibfield  {author} {\bibinfo {author} {\bibfnamefont {G.~N.}\ \bibnamefont
  {Felder}}, \bibinfo {author} {\bibfnamefont {L.}~\bibnamefont {Kofman}}, \
  and\ \bibinfo {author} {\bibfnamefont {A.~D.}\ \bibnamefont {Linde}},\ }\href
  {\doibase 10.1103/PhysRevD.60.103505} {\bibfield  {journal} {\bibinfo
  {journal} {Phys. Rev. D}\ }\textbf {\bibinfo {volume} {60}},\ \bibinfo
  {pages} {103505} (\bibinfo {year} {1999}{\natexlab{b}})},\ \Eprint
  {http://arxiv.org/abs/hep-ph/9903350} {arXiv:hep-ph/9903350} \BibitemShut
  {NoStop}%
\bibitem [{\citenamefont {Haro}(2019)}]{Haro:2018jtb}%
  \BibitemOpen
  \bibfield  {author} {\bibinfo {author} {\bibfnamefont {J.}~\bibnamefont
  {Haro}},\ }\href {\doibase 10.1103/PhysRevD.99.043510} {\bibfield  {journal}
  {\bibinfo  {journal} {Phys. Rev.}\ }\textbf {\bibinfo {volume} {D99}},\
  \bibinfo {pages} {043510} (\bibinfo {year} {2019})},\ \Eprint
  {http://arxiv.org/abs/1807.07367} {arXiv:1807.07367 [gr-qc]} \BibitemShut
  {NoStop}%
\bibitem [{\citenamefont {Dimopoulos}\ \emph {et~al.}(2018)\citenamefont
  {Dimopoulos}, \citenamefont {Donaldson~Wood},\ and\ \citenamefont
  {Owen}}]{Dimopoulos:2017tud}%
  \BibitemOpen
  \bibfield  {author} {\bibinfo {author} {\bibfnamefont {K.}~\bibnamefont
  {Dimopoulos}}, \bibinfo {author} {\bibfnamefont {L.}~\bibnamefont
  {Donaldson~Wood}}, \ and\ \bibinfo {author} {\bibfnamefont {C.}~\bibnamefont
  {Owen}},\ }\href {\doibase 10.1103/PhysRevD.97.063525} {\bibfield  {journal}
  {\bibinfo  {journal} {Phys. Rev. D}\ }\textbf {\bibinfo {volume} {97}},\
  \bibinfo {pages} {063525} (\bibinfo {year} {2018})},\ \Eprint
  {http://arxiv.org/abs/1712.01760} {arXiv:1712.01760 [astro-ph.CO]}
  \BibitemShut {NoStop}%
\bibitem [{\citenamefont {Haro}(2020)}]{Haro:2019ndy}%
  \BibitemOpen
  \bibfield  {author} {\bibinfo {author} {\bibfnamefont {J.}~\bibnamefont
  {Haro}},\ }\href {\doibase 10.1140/epjc/s10052-020-7799-8} {\bibfield
  {journal} {\bibinfo  {journal} {Eur. Phys. J.}\ }\textbf {\bibinfo {volume}
  {C80}},\ \bibinfo {pages} {257} (\bibinfo {year} {2020})},\ \Eprint
  {http://arxiv.org/abs/1904.02393} {arXiv:1904.02393 [gr-qc]} \BibitemShut
  {NoStop}%
\bibitem [{\citenamefont {Wang}\ \emph {et~al.}(1997)\citenamefont {Wang},
  \citenamefont {Mukhanov},\ and\ \citenamefont {Steinhardt}}]{Wang:1997cw}%
  \BibitemOpen
  \bibfield  {author} {\bibinfo {author} {\bibfnamefont {L.-M.}\ \bibnamefont
  {Wang}}, \bibinfo {author} {\bibfnamefont {V.~F.}\ \bibnamefont {Mukhanov}},
  \ and\ \bibinfo {author} {\bibfnamefont {P.~J.}\ \bibnamefont {Steinhardt}},\
  }\href {\doibase 10.1016/S0370-2693(97)01166-0} {\bibfield  {journal}
  {\bibinfo  {journal} {Phys. Lett. B}\ }\textbf {\bibinfo {volume} {414}},\
  \bibinfo {pages} {18} (\bibinfo {year} {1997})},\ \Eprint
  {http://arxiv.org/abs/astro-ph/9709032} {arXiv:astro-ph/9709032} \BibitemShut
  {NoStop}%
\bibitem [{\citenamefont {Martin}\ \emph {et~al.}(2014)\citenamefont {Martin},
  \citenamefont {Ringeval},\ and\ \citenamefont {Vennin}}]{Martin:2013tda}%
  \BibitemOpen
  \bibfield  {author} {\bibinfo {author} {\bibfnamefont {J.}~\bibnamefont
  {Martin}}, \bibinfo {author} {\bibfnamefont {C.}~\bibnamefont {Ringeval}}, \
  and\ \bibinfo {author} {\bibfnamefont {V.}~\bibnamefont {Vennin}},\ }\href
  {\doibase 10.1016/j.dark.2014.01.003} {\bibfield  {journal} {\bibinfo
  {journal} {Phys. Dark Univ.}\ }\textbf {\bibinfo {volume} {5-6}},\ \bibinfo
  {pages} {75} (\bibinfo {year} {2014})},\ \Eprint
  {http://arxiv.org/abs/1303.3787} {arXiv:1303.3787 [astro-ph.CO]} \BibitemShut
  {NoStop}%
\bibitem [{\citenamefont {Jimenez}\ and\ \citenamefont
  {Loeb}(2002)}]{Jimenez:2001gg}%
  \BibitemOpen
  \bibfield  {author} {\bibinfo {author} {\bibfnamefont {R.}~\bibnamefont
  {Jimenez}}\ and\ \bibinfo {author} {\bibfnamefont {A.}~\bibnamefont {Loeb}},\
  }\href {\doibase 10.1086/340549} {\bibfield  {journal} {\bibinfo  {journal}
  {Astrophys. J.}\ }\textbf {\bibinfo {volume} {573}},\ \bibinfo {pages} {37}
  (\bibinfo {year} {2002})},\ \Eprint {http://arxiv.org/abs/astro-ph/0106145}
  {arXiv:astro-ph/0106145 [astro-ph]} \BibitemShut {NoStop}%
\bibitem [{\citenamefont {Moresco}\ \emph
  {et~al.}(2012{\natexlab{a}})\citenamefont {Moresco}, \citenamefont {Verde},
  \citenamefont {Pozzetti}, \citenamefont {Jimenez},\ and\ \citenamefont
  {Cimatti}}]{Moresco:2012by}%
  \BibitemOpen
  \bibfield  {author} {\bibinfo {author} {\bibfnamefont {M.}~\bibnamefont
  {Moresco}}, \bibinfo {author} {\bibfnamefont {L.}~\bibnamefont {Verde}},
  \bibinfo {author} {\bibfnamefont {L.}~\bibnamefont {Pozzetti}}, \bibinfo
  {author} {\bibfnamefont {R.}~\bibnamefont {Jimenez}}, \ and\ \bibinfo
  {author} {\bibfnamefont {A.}~\bibnamefont {Cimatti}},\ }\href {\doibase
  10.1088/1475-7516/2012/07/053} {\bibfield  {journal} {\bibinfo  {journal}
  {JCAP}\ }\textbf {\bibinfo {volume} {1207}},\ \bibinfo {pages} {053}
  (\bibinfo {year} {2012}{\natexlab{a}})},\ \Eprint
  {http://arxiv.org/abs/1201.6658} {arXiv:1201.6658 [astro-ph.CO]} \BibitemShut
  {NoStop}%
\bibitem [{\citenamefont {Moresco}\ \emph
  {et~al.}(2012{\natexlab{b}})\citenamefont {Moresco} \emph
  {et~al.}}]{Moresco:2012jh}%
  \BibitemOpen
  \bibfield  {author} {\bibinfo {author} {\bibfnamefont {M.}~\bibnamefont
  {Moresco}} \emph {et~al.},\ }\href {\doibase 10.1088/1475-7516/2012/08/006}
  {\bibfield  {journal} {\bibinfo  {journal} {JCAP}\ }\textbf {\bibinfo
  {volume} {1208}},\ \bibinfo {pages} {006} (\bibinfo {year}
  {2012}{\natexlab{b}})},\ \Eprint {http://arxiv.org/abs/1201.3609}
  {arXiv:1201.3609 [astro-ph.CO]} \BibitemShut {NoStop}%
\bibitem [{\citenamefont {Moresco}(2015)}]{Moresco:2015cya}%
  \BibitemOpen
  \bibfield  {author} {\bibinfo {author} {\bibfnamefont {M.}~\bibnamefont
  {Moresco}},\ }\href {\doibase 10.1093/mnrasl/slv037} {\bibfield  {journal}
  {\bibinfo  {journal} {Mon. Not. Roy. Astron. Soc.}\ }\textbf {\bibinfo
  {volume} {450}},\ \bibinfo {pages} {L16} (\bibinfo {year} {2015})},\ \Eprint
  {http://arxiv.org/abs/1503.01116} {arXiv:1503.01116 [astro-ph.CO]}
  \BibitemShut {NoStop}%
\bibitem [{\citenamefont {Moresco}\ \emph {et~al.}(2016)\citenamefont
  {Moresco}, \citenamefont {Pozzetti}, \citenamefont {Cimatti}, \citenamefont
  {Jimenez}, \citenamefont {Maraston}, \citenamefont {Verde}, \citenamefont
  {Thomas}, \citenamefont {Citro}, \citenamefont {Tojeiro},\ and\ \citenamefont
  {Wilkinson}}]{Moresco:2016mzx}%
  \BibitemOpen
  \bibfield  {author} {\bibinfo {author} {\bibfnamefont {M.}~\bibnamefont
  {Moresco}}, \bibinfo {author} {\bibfnamefont {L.}~\bibnamefont {Pozzetti}},
  \bibinfo {author} {\bibfnamefont {A.}~\bibnamefont {Cimatti}}, \bibinfo
  {author} {\bibfnamefont {R.}~\bibnamefont {Jimenez}}, \bibinfo {author}
  {\bibfnamefont {C.}~\bibnamefont {Maraston}}, \bibinfo {author}
  {\bibfnamefont {L.}~\bibnamefont {Verde}}, \bibinfo {author} {\bibfnamefont
  {D.}~\bibnamefont {Thomas}}, \bibinfo {author} {\bibfnamefont
  {A.}~\bibnamefont {Citro}}, \bibinfo {author} {\bibfnamefont
  {R.}~\bibnamefont {Tojeiro}}, \ and\ \bibinfo {author} {\bibfnamefont
  {D.}~\bibnamefont {Wilkinson}},\ }\href {\doibase
  10.1088/1475-7516/2016/05/014} {\bibfield  {journal} {\bibinfo  {journal}
  {JCAP}\ }\textbf {\bibinfo {volume} {1605}},\ \bibinfo {pages} {014}
  (\bibinfo {year} {2016})},\ \Eprint {http://arxiv.org/abs/1601.01701}
  {arXiv:1601.01701 [astro-ph.CO]} \BibitemShut {NoStop}%
\bibitem [{\citenamefont {Scolnic}\ \emph {et~al.}(2018)\citenamefont {Scolnic}
  \emph {et~al.}}]{Scolnic:2017caz}%
  \BibitemOpen
  \bibfield  {author} {\bibinfo {author} {\bibfnamefont {D.}~\bibnamefont
  {Scolnic}} \emph {et~al.},\ }\href {\doibase 10.3847/1538-4357/aab9bb}
  {\bibfield  {journal} {\bibinfo  {journal} {Astrophys. J.}\ }\textbf
  {\bibinfo {volume} {859}},\ \bibinfo {pages} {101} (\bibinfo {year}
  {2018})},\ \Eprint {http://arxiv.org/abs/1710.00845} {arXiv:1710.00845
  [astro-ph.CO]} \BibitemShut {NoStop}%
\bibitem [{\citenamefont {Anagnostopoulos}\ \emph {et~al.}(2020)\citenamefont
  {Anagnostopoulos}, \citenamefont {Basilakos},\ and\ \citenamefont
  {Saridakis}}]{Anagnostopoulos:2020ctz}%
  \BibitemOpen
  \bibfield  {author} {\bibinfo {author} {\bibfnamefont {F.~K.}\ \bibnamefont
  {Anagnostopoulos}}, \bibinfo {author} {\bibfnamefont {S.}~\bibnamefont
  {Basilakos}}, \ and\ \bibinfo {author} {\bibfnamefont {E.~N.}\ \bibnamefont
  {Saridakis}},\ }\href {\doibase 10.1140/epjc/s10052-020-8360-5} {\bibfield
  {journal} {\bibinfo  {journal} {Eur. Phys. J.}\ }\textbf {\bibinfo {volume}
  {C80}},\ \bibinfo {pages} {826} (\bibinfo {year} {2020})},\ \Eprint
  {http://arxiv.org/abs/2005.10302} {arXiv:2005.10302 [gr-qc]} \BibitemShut
  {NoStop}%
\bibitem [{\citenamefont {Roberts}\ \emph {et~al.}(2017)\citenamefont
  {Roberts}, \citenamefont {Horne}, \citenamefont {Hodson},\ and\ \citenamefont
  {Leggat}}]{Roberts:2017nkm}%
  \BibitemOpen
  \bibfield  {author} {\bibinfo {author} {\bibfnamefont {C.}~\bibnamefont
  {Roberts}}, \bibinfo {author} {\bibfnamefont {K.}~\bibnamefont {Horne}},
  \bibinfo {author} {\bibfnamefont {A.~O.}\ \bibnamefont {Hodson}}, \ and\
  \bibinfo {author} {\bibfnamefont {A.~D.}\ \bibnamefont {Leggat}},\
  }\href@noop {} {\  (\bibinfo {year} {2017})},\ \Eprint
  {http://arxiv.org/abs/1711.10369} {arXiv:1711.10369 [astro-ph.CO]}
  \BibitemShut {NoStop}%
\bibitem [{\citenamefont {Demianski}\ \emph {et~al.}(2017)\citenamefont
  {Demianski}, \citenamefont {Piedipalumbo}, \citenamefont {Sawant},\ and\
  \citenamefont {Amati}}]{Demianski:2016zxi}%
  \BibitemOpen
  \bibfield  {author} {\bibinfo {author} {\bibfnamefont {M.}~\bibnamefont
  {Demianski}}, \bibinfo {author} {\bibfnamefont {E.}~\bibnamefont
  {Piedipalumbo}}, \bibinfo {author} {\bibfnamefont {D.}~\bibnamefont
  {Sawant}}, \ and\ \bibinfo {author} {\bibfnamefont {L.}~\bibnamefont
  {Amati}},\ }\href {\doibase 10.1051/0004-6361/201628909} {\bibfield
  {journal} {\bibinfo  {journal} {Astron. Astrophys.}\ }\textbf {\bibinfo
  {volume} {598}},\ \bibinfo {pages} {A112} (\bibinfo {year} {2017})},\ \Eprint
  {http://arxiv.org/abs/1610.00854} {arXiv:1610.00854 [astro-ph.CO]}
  \BibitemShut {NoStop}%
\bibitem [{\citenamefont {Benisty}\ and\ \citenamefont
  {Staicova}(2020)}]{Benisty:2020otr}%
  \BibitemOpen
  \bibfield  {author} {\bibinfo {author} {\bibfnamefont {D.}~\bibnamefont
  {Benisty}}\ and\ \bibinfo {author} {\bibfnamefont {D.}~\bibnamefont
  {Staicova}},\ }\href@noop {} {\  (\bibinfo {year} {2020})},\ \Eprint
  {http://arxiv.org/abs/2009.10701} {arXiv:2009.10701 [astro-ph.CO]}
  \BibitemShut {NoStop}%
\bibitem [{\citenamefont {Percival}\ \emph {et~al.}(2010)\citenamefont
  {Percival} \emph {et~al.}}]{Percival:2009xn}%
  \BibitemOpen
  \bibfield  {author} {\bibinfo {author} {\bibfnamefont {W.~J.}\ \bibnamefont
  {Percival}} \emph {et~al.} (\bibinfo {collaboration} {SDSS}),\ }\href
  {\doibase 10.1111/j.1365-2966.2009.15812.x} {\bibfield  {journal} {\bibinfo
  {journal} {Mon. Not. Roy. Astron. Soc.}\ }\textbf {\bibinfo {volume} {401}},\
  \bibinfo {pages} {2148} (\bibinfo {year} {2010})},\ \Eprint
  {http://arxiv.org/abs/0907.1660} {arXiv:0907.1660 [astro-ph.CO]} \BibitemShut
  {NoStop}%
\bibitem [{\citenamefont {Beutler}\ \emph {et~al.}(2011)\citenamefont
  {Beutler}, \citenamefont {Blake}, \citenamefont {Colless}, \citenamefont
  {Jones}, \citenamefont {Staveley-Smith}, \citenamefont {Campbell},
  \citenamefont {Parker}, \citenamefont {Saunders},\ and\ \citenamefont
  {Watson}}]{Beutler:2011hx}%
  \BibitemOpen
  \bibfield  {author} {\bibinfo {author} {\bibfnamefont {F.}~\bibnamefont
  {Beutler}}, \bibinfo {author} {\bibfnamefont {C.}~\bibnamefont {Blake}},
  \bibinfo {author} {\bibfnamefont {M.}~\bibnamefont {Colless}}, \bibinfo
  {author} {\bibfnamefont {D.~H.}\ \bibnamefont {Jones}}, \bibinfo {author}
  {\bibfnamefont {L.}~\bibnamefont {Staveley-Smith}}, \bibinfo {author}
  {\bibfnamefont {L.}~\bibnamefont {Campbell}}, \bibinfo {author}
  {\bibfnamefont {Q.}~\bibnamefont {Parker}}, \bibinfo {author} {\bibfnamefont
  {W.}~\bibnamefont {Saunders}}, \ and\ \bibinfo {author} {\bibfnamefont
  {F.}~\bibnamefont {Watson}},\ }\href {\doibase
  10.1111/j.1365-2966.2011.19250.x} {\bibfield  {journal} {\bibinfo  {journal}
  {Mon. Not. Roy. Astron. Soc.}\ }\textbf {\bibinfo {volume} {416}},\ \bibinfo
  {pages} {3017} (\bibinfo {year} {2011})},\ \Eprint
  {http://arxiv.org/abs/1106.3366} {arXiv:1106.3366 [astro-ph.CO]} \BibitemShut
  {NoStop}%
\bibitem [{\citenamefont {Busca}\ \emph {et~al.}(2013)\citenamefont {Busca}
  \emph {et~al.}}]{Busca:2012bu}%
  \BibitemOpen
  \bibfield  {author} {\bibinfo {author} {\bibfnamefont {N.~G.}\ \bibnamefont
  {Busca}} \emph {et~al.},\ }\href {\doibase 10.1051/0004-6361/201220724}
  {\bibfield  {journal} {\bibinfo  {journal} {Astron. Astrophys.}\ }\textbf
  {\bibinfo {volume} {552}},\ \bibinfo {pages} {A96} (\bibinfo {year}
  {2013})},\ \Eprint {http://arxiv.org/abs/1211.2616} {arXiv:1211.2616
  [astro-ph.CO]} \BibitemShut {NoStop}%
\bibitem [{\citenamefont {Anderson}\ \emph {et~al.}(2013)\citenamefont
  {Anderson} \emph {et~al.}}]{Anderson:2012sa}%
  \BibitemOpen
  \bibfield  {author} {\bibinfo {author} {\bibfnamefont {L.}~\bibnamefont
  {Anderson}} \emph {et~al.},\ }\href {\doibase
  10.1111/j.1365-2966.2012.22066.x} {\bibfield  {journal} {\bibinfo  {journal}
  {Mon. Not. Roy. Astron. Soc.}\ }\textbf {\bibinfo {volume} {427}},\ \bibinfo
  {pages} {3435} (\bibinfo {year} {2013})},\ \Eprint
  {http://arxiv.org/abs/1203.6594} {arXiv:1203.6594 [astro-ph.CO]} \BibitemShut
  {NoStop}%
\bibitem [{\citenamefont {Seo}\ \emph {et~al.}(2012)\citenamefont {Seo} \emph
  {et~al.}}]{Seo:2012xy}%
  \BibitemOpen
  \bibfield  {author} {\bibinfo {author} {\bibfnamefont {H.-J.}\ \bibnamefont
  {Seo}} \emph {et~al.},\ }\href {\doibase 10.1088/0004-637X/761/1/13}
  {\bibfield  {journal} {\bibinfo  {journal} {Astrophys. J.}\ }\textbf
  {\bibinfo {volume} {761}},\ \bibinfo {pages} {13} (\bibinfo {year} {2012})},\
  \Eprint {http://arxiv.org/abs/1201.2172} {arXiv:1201.2172 [astro-ph.CO]}
  \BibitemShut {NoStop}%
\bibitem [{\citenamefont {Ross}\ \emph {et~al.}(2015)\citenamefont {Ross},
  \citenamefont {Samushia}, \citenamefont {Howlett}, \citenamefont {Percival},
  \citenamefont {Burden},\ and\ \citenamefont {Manera}}]{Ross:2014qpa}%
  \BibitemOpen
  \bibfield  {author} {\bibinfo {author} {\bibfnamefont {A.~J.}\ \bibnamefont
  {Ross}}, \bibinfo {author} {\bibfnamefont {L.}~\bibnamefont {Samushia}},
  \bibinfo {author} {\bibfnamefont {C.}~\bibnamefont {Howlett}}, \bibinfo
  {author} {\bibfnamefont {W.~J.}\ \bibnamefont {Percival}}, \bibinfo {author}
  {\bibfnamefont {A.}~\bibnamefont {Burden}}, \ and\ \bibinfo {author}
  {\bibfnamefont {M.}~\bibnamefont {Manera}},\ }\href {\doibase
  10.1093/mnras/stv154} {\bibfield  {journal} {\bibinfo  {journal} {Mon. Not.
  Roy. Astron. Soc.}\ }\textbf {\bibinfo {volume} {449}},\ \bibinfo {pages}
  {835} (\bibinfo {year} {2015})},\ \Eprint {http://arxiv.org/abs/1409.3242}
  {arXiv:1409.3242 [astro-ph.CO]} \BibitemShut {NoStop}%
\bibitem [{\citenamefont {Tojeiro}\ \emph {et~al.}(2014)\citenamefont {Tojeiro}
  \emph {et~al.}}]{Tojeiro:2014eea}%
  \BibitemOpen
  \bibfield  {author} {\bibinfo {author} {\bibfnamefont {R.}~\bibnamefont
  {Tojeiro}} \emph {et~al.},\ }\href {\doibase 10.1093/mnras/stu371} {\bibfield
   {journal} {\bibinfo  {journal} {Mon. Not. Roy. Astron. Soc.}\ }\textbf
  {\bibinfo {volume} {440}},\ \bibinfo {pages} {2222} (\bibinfo {year}
  {2014})},\ \Eprint {http://arxiv.org/abs/1401.1768} {arXiv:1401.1768
  [astro-ph.CO]} \BibitemShut {NoStop}%
\bibitem [{\citenamefont {Bautista}\ \emph {et~al.}(2018)\citenamefont
  {Bautista} \emph {et~al.}}]{Bautista:2017wwp}%
  \BibitemOpen
  \bibfield  {author} {\bibinfo {author} {\bibfnamefont {J.~E.}\ \bibnamefont
  {Bautista}} \emph {et~al.},\ }\href {\doibase 10.3847/1538-4357/aacea5}
  {\bibfield  {journal} {\bibinfo  {journal} {Astrophys. J.}\ }\textbf
  {\bibinfo {volume} {863}},\ \bibinfo {pages} {110} (\bibinfo {year}
  {2018})},\ \Eprint {http://arxiv.org/abs/1712.08064} {arXiv:1712.08064
  [astro-ph.CO]} \BibitemShut {NoStop}%
\bibitem [{\citenamefont {de~Carvalho}\ \emph {et~al.}(2018)\citenamefont
  {de~Carvalho}, \citenamefont {Bernui}, \citenamefont {Carvalho},
  \citenamefont {Novaes},\ and\ \citenamefont {Xavier}}]{deCarvalho:2017xye}%
  \BibitemOpen
  \bibfield  {author} {\bibinfo {author} {\bibfnamefont {E.}~\bibnamefont
  {de~Carvalho}}, \bibinfo {author} {\bibfnamefont {A.}~\bibnamefont {Bernui}},
  \bibinfo {author} {\bibfnamefont {G.~C.}\ \bibnamefont {Carvalho}}, \bibinfo
  {author} {\bibfnamefont {C.~P.}\ \bibnamefont {Novaes}}, \ and\ \bibinfo
  {author} {\bibfnamefont {H.~S.}\ \bibnamefont {Xavier}},\ }\href {\doibase
  10.1088/1475-7516/2018/04/064} {\bibfield  {journal} {\bibinfo  {journal}
  {JCAP}\ }\textbf {\bibinfo {volume} {1804}},\ \bibinfo {pages} {064}
  (\bibinfo {year} {2018})},\ \Eprint {http://arxiv.org/abs/1709.00113}
  {arXiv:1709.00113 [astro-ph.CO]} \BibitemShut {NoStop}%
\bibitem [{\citenamefont {Ata}\ \emph {et~al.}(2018)\citenamefont {Ata} \emph
  {et~al.}}]{Ata:2017dya}%
  \BibitemOpen
  \bibfield  {author} {\bibinfo {author} {\bibfnamefont {M.}~\bibnamefont
  {Ata}} \emph {et~al.},\ }\href {\doibase 10.1093/mnras/stx2630} {\bibfield
  {journal} {\bibinfo  {journal} {Mon. Not. Roy. Astron. Soc.}\ }\textbf
  {\bibinfo {volume} {473}},\ \bibinfo {pages} {4773} (\bibinfo {year}
  {2018})},\ \Eprint {http://arxiv.org/abs/1705.06373} {arXiv:1705.06373
  [astro-ph.CO]} \BibitemShut {NoStop}%
\bibitem [{\citenamefont {Abbott}\ \emph {et~al.}(2019)\citenamefont {Abbott}
  \emph {et~al.}}]{Abbott:2017wcz}%
  \BibitemOpen
  \bibfield  {author} {\bibinfo {author} {\bibfnamefont {T.~M.~C.}\
  \bibnamefont {Abbott}} \emph {et~al.} (\bibinfo {collaboration} {DES}),\
  }\href {\doibase 10.1093/mnras/sty3351} {\bibfield  {journal} {\bibinfo
  {journal} {Mon. Not. Roy. Astron. Soc.}\ }\textbf {\bibinfo {volume} {483}},\
  \bibinfo {pages} {4866} (\bibinfo {year} {2019})},\ \Eprint
  {http://arxiv.org/abs/1712.06209} {arXiv:1712.06209 [astro-ph.CO]}
  \BibitemShut {NoStop}%
\bibitem [{\citenamefont {Molavi}\ and\ \citenamefont
  {Khodam-Mohammadi}(2019)}]{Molavi:2019mlh}%
  \BibitemOpen
  \bibfield  {author} {\bibinfo {author} {\bibfnamefont {Z.}~\bibnamefont
  {Molavi}}\ and\ \bibinfo {author} {\bibfnamefont {A.}~\bibnamefont
  {Khodam-Mohammadi}},\ }\href {\doibase 10.1140/epjp/i2019-12723-x} {\bibfield
   {journal} {\bibinfo  {journal} {Eur. Phys. J. Plus}\ }\textbf {\bibinfo
  {volume} {134}},\ \bibinfo {pages} {254} (\bibinfo {year} {2019})},\ \Eprint
  {http://arxiv.org/abs/1906.05668} {arXiv:1906.05668 [gr-qc]} \BibitemShut
  {NoStop}%
\bibitem [{\citenamefont {Hogg}\ \emph {et~al.}(2020)\citenamefont {Hogg},
  \citenamefont {Martinelli},\ and\ \citenamefont {Nesseris}}]{Hogg:2020ktc}%
  \BibitemOpen
  \bibfield  {author} {\bibinfo {author} {\bibfnamefont {N.~B.}\ \bibnamefont
  {Hogg}}, \bibinfo {author} {\bibfnamefont {M.}~\bibnamefont {Martinelli}}, \
  and\ \bibinfo {author} {\bibfnamefont {S.}~\bibnamefont {Nesseris}},\
  }\href@noop {} {\  (\bibinfo {year} {2020})},\ \Eprint
  {http://arxiv.org/abs/2007.14335} {arXiv:2007.14335 [astro-ph.CO]}
  \BibitemShut {NoStop}%
\bibitem [{\citenamefont {Martinelli}\ \emph {et~al.}(2020)\citenamefont
  {Martinelli} \emph {et~al.}}]{Martinelli:2020hud}%
  \BibitemOpen
  \bibfield  {author} {\bibinfo {author} {\bibfnamefont {M.}~\bibnamefont
  {Martinelli}} \emph {et~al.} (\bibinfo {collaboration} {EUCLID}),\
  }\href@noop {} {\  (\bibinfo {year} {2020})},\ \Eprint
  {http://arxiv.org/abs/2007.16153} {arXiv:2007.16153 [astro-ph.CO]}
  \BibitemShut {NoStop}%
\bibitem [{\citenamefont {Chen}\ \emph {et~al.}(2019)\citenamefont {Chen},
  \citenamefont {Huang},\ and\ \citenamefont {Wang}}]{Chen:2018dbv}%
  \BibitemOpen
  \bibfield  {author} {\bibinfo {author} {\bibfnamefont {L.}~\bibnamefont
  {Chen}}, \bibinfo {author} {\bibfnamefont {Q.-G.}\ \bibnamefont {Huang}}, \
  and\ \bibinfo {author} {\bibfnamefont {K.}~\bibnamefont {Wang}},\ }\href
  {\doibase 10.1088/1475-7516/2019/02/028} {\bibfield  {journal} {\bibinfo
  {journal} {JCAP}\ }\textbf {\bibinfo {volume} {02}},\ \bibinfo {pages} {028}
  (\bibinfo {year} {2019})},\ \Eprint {http://arxiv.org/abs/1808.05724}
  {arXiv:1808.05724 [astro-ph.CO]} \BibitemShut {NoStop}%
\bibitem [{\citenamefont {Handley}\ \emph {et~al.}(2015)\citenamefont
  {Handley}, \citenamefont {Hobson},\ and\ \citenamefont
  {Lasenby}}]{Handley:2015fda}%
  \BibitemOpen
  \bibfield  {author} {\bibinfo {author} {\bibfnamefont {W.~J.}\ \bibnamefont
  {Handley}}, \bibinfo {author} {\bibfnamefont {M.~P.}\ \bibnamefont {Hobson}},
  \ and\ \bibinfo {author} {\bibfnamefont {A.~N.}\ \bibnamefont {Lasenby}},\
  }\href {\doibase 10.1093/mnrasl/slv047} {\bibfield  {journal} {\bibinfo
  {journal} {Mon. Not. Roy. Astron. Soc.}\ }\textbf {\bibinfo {volume} {450}},\
  \bibinfo {pages} {L61} (\bibinfo {year} {2015})},\ \Eprint
  {http://arxiv.org/abs/1502.01856} {arXiv:1502.01856 [astro-ph.CO]}
  \BibitemShut {NoStop}%
\bibitem [{\citenamefont {Lewis}(2019)}]{Lewis:2019xzd}%
  \BibitemOpen
  \bibfield  {author} {\bibinfo {author} {\bibfnamefont {A.}~\bibnamefont
  {Lewis}},\ }\href@noop {} {\  (\bibinfo {year} {2019})},\ \Eprint
  {http://arxiv.org/abs/1910.13970} {arXiv:1910.13970 [astro-ph.IM]}
  \BibitemShut {NoStop}%
\bibitem [{\citenamefont {Aresté~Saló}\ \emph {et~al.}(2021)\citenamefont
  {Aresté~Saló}, \citenamefont {Benisty}, \citenamefont {Guendelman},\ and\
  \citenamefont {de~Haro}}]{AresteSalo:2021wgb}%
  \BibitemOpen
  \bibfield  {author} {\bibinfo {author} {\bibfnamefont {L.}~\bibnamefont
  {Aresté~Saló}}, \bibinfo {author} {\bibfnamefont {D.}~\bibnamefont
  {Benisty}}, \bibinfo {author} {\bibfnamefont {E.~I.}\ \bibnamefont
  {Guendelman}}, \ and\ \bibinfo {author} {\bibfnamefont {J.}~\bibnamefont
  {de~Haro}},\ }\href@noop {} {\  (\bibinfo {year} {2021})},\ \Eprint
  {http://arxiv.org/abs/2103.07892} {arXiv:2103.07892 [astro-ph.CO]}
  \BibitemShut {NoStop}%
\end{thebibliography}%
\end{document}